# Postoperative brain volumes are associated with one-year neurodevelopmental outcome in children with severe congenital heart disease

Eliane Meuwly[1]** and Maria Feldmann[1]**, Walter Knirsch[2,3], Michael von Rhein[1], Kelly Payette[4], Hitendu Dave[5], Ruth O'Gorman Tuura[3,4], Raimund Kottke[6], Cornelia Hagmann[7], Beatrice Latal[1, 3]† and András Jakab[4]†*, and the Research Group Heart and Brain, **shared first authorship, †shared last authorship,
**Corresponding author:** András Jakab, Email: Andras.Jakab@kispi.uzh.ch

Children with congenital heart disease (CHD) remain at risk for neurodevelopmental impairment despite improved peri- and intraoperative care. Our prospective cohort study aimed to determine the relationship between perioperative brain volumes and neurodevelopmental outcome in neonates with severe CHD. Pre- and postoperative cerebral MRI was acquired in term born neonates with CHD undergoing neonatal cardiopulmonary bypass surgery. Brain volumes were measured using an atlas prior based automated method. One-year outcome was assessed with the Bayley-III. CHD infants (n=77) had lower pre- and postoperative total and regional brain volumes compared to controls (n=44, all p<0.01). CHD infants had poorer cognitive and motor outcome (p≤0.0001) and a trend towards lower language composite score compared to controls (p=0.06). The total and selected regional postoperative brain volumes predicted cognitive and language outcome (all p<0.04). This association was independent of length of intensive care unit stay for total, cortical, temporal, frontal and cerebellar volumes. In CHD neonates undergoing cardiopulmonary bypass surgery, pre- and postoperative brain volumes are reduced, and postoperative brain volumes predict cognitive and language outcome at one year. Reduced cerebral volumes in CHD patients could serve as a biomarker for impaired outcome.

## Background

Congenital heart disease (CHD) is the most common congenital malformation in childhood[1]. Despite dramatic improvements in peri- and intraoperative care that have led to an increase in survival rates, children with severe CHD remain at risk for neurodevelopmental impairment [1,2]. Cognitive, language and motor functions may be impaired during early childhood [3-5], but impairments often persist and include deficits in visuomotor and executive functions in adolescence [6,7]. Cerebral magnetic resonance imaging (MRI) studies have shown a high incidence of usually small pre- and postoperative white matter injuries (WMI) [2,10] and strokes [2,11,12] in neonates with CHD, and delayed brain maturation and growth has also been described in this population [13-16]. Volumetric analyses of neonatal cerebral MRI examinations have revealed that infants with severe CHD have smaller brain volumes compared to healthy born infants [8,17], with all brain regions being equally affected [17]. There is mounting evidence that delayed brain maturation and brain injuries in neonates with CHD are associated with poorer neurodevelopmental outcome. Recently published studies have reported that new postoperative cerebral lesions on MRI correlate with lower cognitive and language outcome at one year [18]. The white matter seems particularly predictive of early childhood motor [12] and school-age cognitive outcome [19]. Moreover, delayed brain maturation has been shown to predict 2-year neurodevelopmental outcome [11]. While total brain volume reduction has been shown to correlate with neurobehaviour prior to neonatal surgery (4), the role of regional brain volumes in predicting neurodevelopmental outcome in early childhood is yet unknown. Elucidating the association between regional brain volume reduction and specific neurodevelopmental outcomes may enable us to determine the antecedents of later altered development [17].

The primary aim of our study was to examine the predictive value of pre- and postoperative total and regional brain volumes for one-year neurodevelopmental outcome in children with severe CHD, and to investigate if such an effect is independent of patient-specific (e.g. sex, severity of CHD, socioeconomic status (SES)) and perioperative risk factors. A secondary aim was to examine brain growth between pre- and postoperative MRI in infants with CHD in comparison to cross-sectional brain volumes from healthy infants, in order to assess if brain growth is delayed in children with CHD, and whether the delay is predictive of neurodevelopmental outcome. We hypothesized that lower regional brain volumes would be associated with poorer neurodevelopmental outcome independent of patient-specific and perioperative risk factors. We further hypothesized that brain growth would be delayed in CHD infants, and that the delay would be predictive of poorer outcome.

## Methods

***Patient population.*** This study is part of an ongoing prospective cohort study investigating neurodevelopmental outcome and the timing of cerebral injuries in infants operated for CHD [17]. Between December 2009 and August 2016, term born neonates (> 36 weeks gestation) with severe CHD undergoing cardiopulmonary bypass surgery (CPB) were consecutively enrolled in the study. Written informed consent was provided from the parents or legal guardians and the study was approved by the regional ethics committee of the Canton Zurich (Kantonale Ethikkommission Zürich). The study was carried out in accordance with the principles enunciated in the Declaration of Helsinki and the guidelines of Good Clinical Practice. Neonates were enrolled in the study after being admitted to the paediatric cardiac intensive care unit (ICU). Neonates with a suspected or confirmed genetic disorder or syndrome were excluded. Infants included in the study

[1] Child Development Centre, University Children's Hospital, Zurich, Switzerland
[2] Paediatric Cardiology, Paediatric Heart Centre, Department of Surgery, University Children's Hospital Zurich, Switzerland
[3] Children's Research Center, University Children's Hospital Zurich, Switzerland
[4] Centre for MR Research, University Children's Hospital of Zurich, Switzerland
[5] Paediatric Cardiovascular Surgery, Paediatric Heart Centre, Department of Surgery, University Children's Hospital Zurich, Switzerland
[6] Department of Diagnostic Imaging, University Children's Hospital Zurich, Switzerland
[7] Department of Neonatology and Pediatric Intensive Care, University Children's Hospital Zurich, Switzerland

underwent cerebral MRI before and after surgery and a neurodevelopmental assessment at 12 months of age, using the Bayley Scales of Infant and Toddler Development, Third Edition (Bayley-III). The Bayley–III assesses different developmental domains and provides three main composite scores: the cognitive composite score (CCS), the language composite score (LCS) and the motor composite score (MCS) with a mean score of 100 and a standard deviation of +/- 15. The Bayley-III was administered by trained developmental paediatricians. SES was estimated based on maternal education and paternal occupation [20] yielding a score ranging from 2-12 with higher scores indicating higher SES.

Healthy term born infants were recruited as controls from the well-baby maternity unit of the University Hospital Zurich. None of the control infants had any cerebral lesions on MRI, and none of the controls had been admitted to the neonatal unit.

***Surgical procedure.*** Cardiac surgery included biventricular repair by arterial switch or Rastelli operation for patients with dextro-transposition of the great arteries (dTGA), complex aortic arch surgery, aortopulmonary shunt procedure and early Fallot repair. As univentricular palliation Norwood-type stage I palliation for patients with hypoplastic left heart syndrome and other forms of univentricular physiology with a hypoplastic aortic arch was performed [17].

***Cerebral MRI and image post processing.*** Neonatal cerebral MRI was performed during natural sleep on a 3.0 T clinical MRI scanner using an 8-channel head coil (GE Signa MR750). Ear plugs (attenuation: 24 dB; Earsoft; Aearo) and Minimuffs (attenuation: 7 dB; Natus) were applied for noise protection. Oxygen saturation was monitored during scanning, and a neonatologist and a neonatal nurse were present during the MRI investigation. 2D fast spin-echo T2-weighted anatomical sequences were acquired in axial, sagittal and coronal planes. The sequence parameters for the anatomical MRI were the following. TE/TR: 97/5900 ms, flip angle: 90°, number of averages: 2, acquisition matrix: 512 * 320, in-plane voxel dimensions: 0.7 * 0.7 mm resampled to 0.35 * 0.35 mm, slice thickness: 2.5 mm, slice gap: 0.2 mm. The anatomical MRI was followed by diffusion tensor imaging, proton MR spectroscopy and arterial spin labelling sequences. Further details about the imaging setting and MRI sequence parameters for this ongoing study have been reported previously [10].

Each MRI scan was scored for the presence of brain abnormalities by a trained neuroradiologist blinded to all clinical characteristics [10]. During the course of the study, a scanner software update was performed. Therefore, the cohort was divided into two groups (MRI Cohort 1 and 2) according to the scanner upgrade status, and this variable was included in the statistical tests as a covariate.

An automatic image analysis workflow was utilized to measure total and regional brain volumes in each neonate. First, T2-weighted anatomical images were re-sampled to a super-resolution image using the principles described in previous reports [21]. Non-brain tissue parts of the axial, coronal and sagittal T2-weighted images were removed by applying image masks generated from an age-specific neonatal neuroanatomical atlas. The masked, axial and coronal images were co-registered to the sagittal image using a mutual information based affine and non-linear registration as implemented in the Niftireg image registration package [22]. A slice-wise reconstruction step, which is usually employed during super-resolution reconstructions [23], was skipped due to the relative scarcity of subject motion, and since any motion-corrupted scans had been repeated. Bias field correction of the images was performed by the N4ITK filter in the Slicer 3D software [24], while further steps of the re-sampling were carried out using the BTK Toolkit [25]. The original, three-plane T2 images and the 3DT2 reconstruction are illustrated in Figure 1/a.

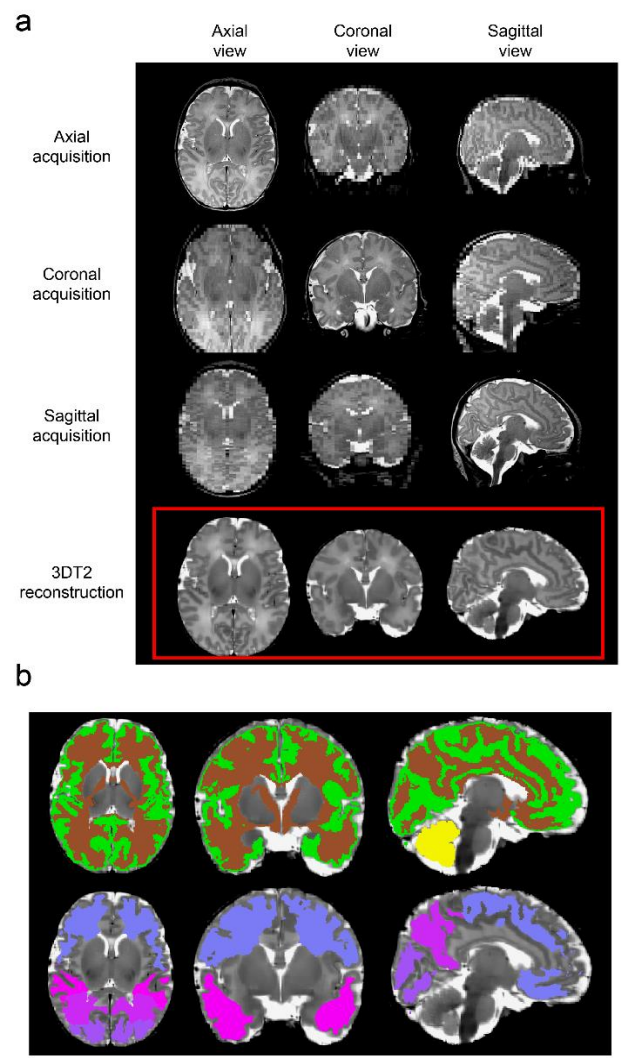

**Figure 1. 3D reconstruction and brain segmentation.** *a. reconstruction of 3D T2-weighted images (bottom row) from three orthogonally acquired images, b. example segmentations: cortex, white matter and cerebellum (top row), and four-lobe segmentation (bottom row).*

Anatomical priors from the gestational age-specific neonatal anatomical atlas ALBERTs [26] were matched to the subjects reconstructed images using a freeform non-linear deformation (reg_f3d command in the Niftireg software) with a fine transformation grid of 8 x 8 x 8 mm, penalty term was bending energy with a setting of 0.0035 to allow for more accurate alignment between the template and subject's images. The anatomical masks of white matter, grey matter (cortex only) and lobar subdivisions of the supratentorial white matter (frontal, parietal, temporal and occipital lobe) and cerebellum were matched to the T2-weighted super-resolution images. Total brain (parenchymal) volume was estimated by constructing a whole-brain mask, which included the diencephalon, cortical mantle, supratentorial white matter, but not the intra- and extra axial cerebrospinal fluid spaces or cerebellum. The patient-specific priors [27] were thresholded at 50 % probability, and their volume was stored for the statistical evaluations (example segmentations: Figure 1/b).

To validate the automated atlas-based volumetry workflow against manual volumetry, a stereological analysis of a sub-set of the study sample was performed. Stereology was carried out using the EasyMeasure software package [28], following a method that has previously been described in detail elsewhere [29]. The validation showed good agreement between the two methods, with a Pearson correlation coefficient ranging from 0.930 (p < 0.01) for the total brain volume and 0.827 (p < 0.01) for the cerebellar volume. In-depth details and results of the validation process are given in the Supplementary Information.

Following this initial validation, all subsequent analyses including brain volumes were performed by the automated pipeline.

Brain growth was defined as the weekly rate of volumetric change of the brain and its major subdivisions, calculated by using the volumetric measurements of the pre- and postoperatively acquired MRI scans.

***Statistical analyses.*** All analyses were conducted using the statistical software R, Version 3.5.0 [30]. Results were considered to be significant at a p-level of < 0.05. Subject demographics are listed in table 1 as mean and standard deviation (SD) for normally distributed continuous variables and median and interquartile range (IQR) for non-normally distributed continuous variables. Categorical variables are reported as proportions. Groupwise comparisons for continuous variables were performed by t tests and Man-Whitney U tests as appropriate for the sample distributions and comparisons for categorical variables were assessed using Fisher´s exact test. To quantify the associations between brain volumes and neurodevelopmental outcome we calculated a series of multiple linear regression models adjusted for postmenstrual age at MRI, SES, sex and MRI cohort, with the latter accounting for a software update on the scanner during the time of the study. To further correct for possible confounds associated with the length of ICU stay, this variable was consecutively introduced into the model. Square root transformation for dependent variables such as brain volumes and Bayley-III composite scores was performed to meet the assumptions of multiple linear regression models. Pearson correlation coefficients were calculated to determine the relationship between potential risk factors and total brain volumes in infants with CHD.

## Results

***Study population.*** In this prospective cohort study, 77 infants with a severe CHD were enrolled between December 2009 and August 2016 and 44 healthy infants were recruited as controls. In the CHD group, 70 had a preoperative MRI, 68 had a postoperative MRI and 60 had both a pre-and postoperative MRI. Reasons for missing preoperative brain MRI included patient instability and too short of a time period between recruitment and surgery (n = 7). In addition, 3 scans could not be included in the regional volumetric analyses due to motion artefacts. Postoperatively, 10 patients did not have a cerebral MRI because infants were too old to undergo MRI in natural sleep (n = 6), or for logistical reasons. Furthermore, 6 postoperative scans could not be analysed due to poor image quality (see Supplementary Fig. S1). Among the MRIs of healthy controls, one could not be analysed due to motion artefacts. Hence, volumetric analysis was performed for 67 CHD infants preoperatively, 62 postoperatively and for 43 controls. Mean (± SD) time between pre-and postoperative MRI scans in CHD infants was 20 (± 8) days.

***Table 1. Characteristics of infants with CHD and healthy controls.*** *M = Median, m = mean, IQR = interquartile range, SD = standard deviation. CHD, congenital heart disease; GA, gestational age; ICU, intensive care unit; SES, socioeconomic status; PMA, postmenstrual age; MRI, magnetic resonance imaging; NA, not applicable. Groupwise comparison for categorical variables was performed by Fisher´s exact test and for continuous variables t test and Man-Whitney U test were applied appropriate for sample distribution.*

| | Infants with CHD | Controls | *p* |
|---|---|---|---|
| n | 77 | 44 | |
| Male, n (%) | 58 (75.3) | 20 (45.5) | 0.002 |
| Birthweight, g (m (SD)) | 3347.7 (491.5) | 3432.0 (402.5) | 0.34 |
| GA, wks (m (SD)) | 39.4 (1.3) | 39.6 (1.2) | 0.42 |
| Head circumference at birth, cm (m (SD)) | 34.6 (1.3) | 35.1 (1.2) | 0.035 |
| Apgar 5 (M [IQR]) | 9.0 [8.0, 9.0] | 9.0 [9.0, 9.0] | 0.29 |
| SES (2-12) (M [IQR]) | 8.0 [7.0, 10.0] | 12.0 [10.0, 12.0] | <0.001 |
| Cyanotic vitium, n (%) | 69 (89.6) | NA | |
| Highest lactate preoperatively, mmol/l (M [IQR]) | 4.3 [3.1, 6.1] | NA | |
| Age at surgery, d (M [IQR]) | 12.0 [9.0, 16.0] | NA | |
| Extracorporeal circulation time, min (m (SD)) | 153.1 (78.0) | NA | |
| Aortic cross clamp time, min (m (SD)) | 93.4 (52.7) | NA | |
| Lowest temperature on bypass, °C (M [IQR]) | 28.1 [23.1, 31.5] | NA | |
| Days on ICU postoperatively (m (SD)) | 7.9 (10.1) | NA | |
| Age, days (M [IQR]) at | | | |
| Preoperative MRI | 7.0 [5.2, 8.8] | 21.0 [16.0, 28.0] | |
| Postoperative MRI | 25.0 [20.8, 31.0] | | |
| PMA, weeks (M [IQR]) at | | | |
| Preoperative MRI | 40.2 [39.3, 41.2] | 42.3 [41.2, 43.6] | |
| Postoperative MRI | 43.2 [41.4, 44.1] | | |

**Table 1** presents demographic characteristics of the CHD infants and controls. Complete data (pre- and postoperative analysed brain scans and Bayley-III assessment) were collected for a total of 50 CHD infants. Details on CHD diagnoses are listed in **Table 2**.

***Cerebral MRI: pre- and postoperative brain injury.*** On the preoperatively acquired MRI scans, 14 out of 67 (20.9 %) patients had a brain lesion, comprised of either white matter injury (WMI) and/or stroke. WMI was detected in 14 (19.7 %) (11 minimal, 2 moderate and 1 severe injury), and stroke was detected in 4 (8.5 %) infants (1 anterior cerebral artery, 2 middle cerebral artery, 1 posterior cerebral artery). Two infants had both WMI and stroke. On the postoperative MRI new postoperative lesions were found in 2 of the 68 scanned infants (2.9 %) and all were minimal white matter injuries. In 13 of the 16 infants with preoperative stroke and/or WMI, lesions persisted to postoperative scans (81.3 %).

***Cerebral MRI: volumetric findings.*** We report three main findings regarding the differences in cerebral volumes between CHD infants and controls.

*Table 2. Distribution of cardiac diagnoses of infants with CHD. CHD, congenital heart disease; dTGA, dextro-transposition of the great arteries; DORV, double outlet right ventricle, HLHS, hypoplastic left heart syndrome; HLHC, hypoplastic left heart complex.*

| Cardiac diagnosis | n |
|---|---|
| Biventricular | 62 |
| dTGA | 38 |
| Aortic arch anomaly | 8 |
| DORV TGA type | 7 |
| Other† | 9 |
| Univentricular | 15 |
| HLHS/HLHC | 9 |
| Other* | 6 |

*† Other biventricular diagnoses include neonates with atrioventricular septal defect (n = 2), ventricular septal defect (n = 1), pulmonary atresia with ventricular septal defect (n = 4), tetralogy of Fallot (n = 1) and levo-transposition of the great arteries (n = 1).*

*\* Other univentricular diagnoses (non HLHS/HLHC) include neonates with tricuspid atresia (n = 3), pulmonary atresia with intact ventricular septum (n = 1), double inlet left ventricle (n = 1) and heterotaxy syndrome (n = 1).*

*Table 3. Pre- and postoperatively measured absolute brain volumes of CHD infants and controls in cm3 given as mean (standard deviation). P-values and ß coefficients (ß) based on multiple linear regression analyses adjusted for sex, postmenstrual age at MRI, MRI cohort.*

| | Brain Volume | | | Group comparison | |
|---|---|---|---|---|---|
| | Preop n = 67 | Postop n = 62 | Controls n = 43 | Preop vs. Controls, ß | Postop vs. Controls, ß |
| Total brain | 371.3 (51.5) | 395.2 (40.8) | 434.7 (51.6) | -1.21† | -1.43† |
| Cortex | 253.8 (43.5) | 273.8 (33.7) | 302.5 (39.4) | -1.06† | -1.29† |
| White matter | 189.4 (26.7) | 196.2 (20.7) | 216.6 (26) | -0.81† | -0.97† |
| Frontal lobe | 105.7 (13.4) | 111.5 (11.3) | 124.3 (14.9) | -0.72† | -0.82† |
| Parietal lobe | 69.8 (11.3) | 73.1 (8.6) | 80.4 (11.8) | -0.45** | -0.59† |
| Occipital lobe | 33.1 (6.9) | 35.7 (4.1) | 39.6 (4.7) | -0.43** | -0.44† |
| Temporal lobe | 62.8 (8.8) | 66.6 (7.2) | 72.2 (8.8) | -0.46† | -0.51† |
| Cerebellum | 31.1 (5.8) | 33.9 (4.7) | 37.7 (5.9) | -0.36** | -0.46† |

*P-values coded as: \*\* = p < 0.001, † = p < 0.0001. Preop, preoperative; Postop, postoperative.*

First, CHD infants had lower total and regional brain volumes in all regions pre- and postoperatively without a specific regional predilection (Table 3). When comparing relative regional brain volumes (corrected for total brain volume) of CHD infants with controls we found the preoperative relative frontal lobe to be smaller in the CHD group (ß = -0.18 95 % CI -0.32 to -0.04, p = 0.012), whereas the preoperative relative parietal lobe appeared to be larger (ß = 0.12, 95 % CI 0.01 to 0.23, p = 0.035). For postoperative relative regional brain volumes, no difference was found. Second, CHD infants with a preoperative lesion had significantly smaller cerebellar volumes on the preoperative MRI (28.14 cm³ vs. 31.85 cm³, 95 % CI -7.07 to -0.35, p = 0.031), but not on the postoperative MRI (32.24 cm3 vs. 34.7 cm³, 95 % CI -5.38 to 0.45, p = 0.095). There was no significant difference in total brain volume (neither pre-, nor postoperative) or in other regional brain volumes between patients with and without preoperative brain lesions. Third, total brain volume was found to be smaller in female compared to male CHD infants (preoperative: mean 347.27 cm³ vs. 379.53 cm³ (95 % CI 4.26 to 60.25, p = 0.025); postoperative: 369.12 cm³ vs. 404.29 cm³ (95% CI 13.05 to 57.28, p = 0.002). In contrast, there was no difference in total brain volume between male and female controls (95 % CI -50.13 to 13.35, p = 0.249).

Preoperative and postoperative total brain volumes correlated with head circumference at birth (preoperative: r = 0.43, 95 % CI 0.20 to 0.62, p = 0.0005; postoperative: r = 0.65, 95 % CI 0.46 to 0.78, p < 0.0001), gestational age (preoperative: r = 0.29, 95 % CI 0.05 to 0.5, p = 0.019; postoperative: r = 0.3, 95 % CI 0.05 to 0.5, p = 0.018) and birth weight (preoperative: r = 0.47, 95 % CI 0.25 to 0.63, p < 0.0001, postoperative: r = 0.57, 95 % CI 0.37 to 0.71, p < 0.0001). Furthermore, we investigated the impact of perinatal and intraoperative variables (i. e. 5 min Apgar score, extracorporeal circulation time, lowest temperature on bypass, aortic cross clamp time) on postoperative total brain volumes and found no significant correlation (see Supplementary Table S1 online). However, a longer length of ICU stay was associated with smaller postoperative total brain volume (r = -0.25, 95 % CI -0.48 to -0.00, p = 0.047). Group comparison further revealed no difference in postoperative whole brain volume for infants with or without dTGA (mean 393.38 cm³ vs. 396.63 cm³, 95 % CI -24.33 to 17.83, p = 0.53) or univentricular heart defect (396.47 cm³ vs. 388.66 cm³, 95 % CI -20.56 to 36.19, p = 0.76).

***Neurodevelopmental outcome at one year.*** Neurodevelopmental assessments were performed at a median [IQR] age of 12.0 months [12.0, 13.0] for CHD infants and 12.0 months [12.0, 12.0] for controls. One infant died prior to the one-year examination and 2 of 76 surviving CHD infants did not return for the neurodevelopmental assessment (follow up rate 97.4 %). In two infants, the language score could not be assessed due to non-German language. Among the 44 control children 7 were lost to follow up (follow up rate 84.1 %). The Bayley-III CCS and MCS scores were significantly lower in CHD patients than in controls: CCS (mean (SD)) 105 (15.3) vs. 117.1 (11.3), (95% CI 6.43 to 17.68, p<0.0001); MCS 92.7 (15.4) vs. 103.8 (10.9), (95 % CI 5.50 to 16.74, p = 0.0002). For the LCS only a trend towards lower scores in CHD patients was found: 92.7 (14.0) vs. 97.5 (9.9), (95 % CI -0.30 to 9.96, p = 0.06). After adjustment for SES, the differences in CCS (ß = -0.34, 95 % CI -0.66 to -0.02, p = 0.037) and MCS (ß = -0.37, 95 % CI -0.72 to -0.03, p = 0.036) among infants with CHD compared to controls remained, while no association was found for the LCS (ß = -0.06, 95 % CI -0.37 to 0.25, p = 0.70). Multiple linear regression, adjusted for SES revealed that MCS was higher for CHD patients with dTGA compared to patients with non-dTGA diagnoses (ß = 0.50, 95 % CI 0.13 to 0.88, p = 0.009), while CCS and LCS was not (CCS ß = 0.23, 95 % CI -0.12 to 0.58, p = 0.19; LCS ß = -0.24, 95 % CI -0.58 to 0.09, p = 0.15). Length of ICU stay was negatively associated with CCS (ß = -0.018, 95 % CI -0.04 to -0.00, p = 0.045) and MCS (ß = -0.02, 95 % CI -0.04 to -0.00,

p = 0.036), but not with LCS (ß = -0.003, 95 % CI -0.02 to 0.01, p = 0.73).

**Brain growth.** All postoperative brain volumes in infants with CHD were significantly larger than preoperative brain volumes (see Supplementary Table S2 online). We furthermore evaluated whether the brain growth rate was lower in CHD infants than the estimated growth from the cross-sectionally acquired growth in the healthy controls. The estimated weekly growth of the total brain volume of controls, estimated by fitting a linear function to the cross sectional volumetric data of controls, exceeded that of the CHD infants (controls 22.3 cm$^3$ per week, CHD 11.0 cm$^3$ per week). The estimated cerebral growth rate in controls exceeded the observed growth rate of 91.4 % of the CHD infants (**Figure 2, Supplementary Figure S4, Supplementary Figure S5**).

**Relationship between perioperative cerebral MRI findings with neurodevelopmental outcome.** Cerebral lesions on pre- and postoperative MRI did not predict neurodevelopmental outcome. *Preoperative* brain volumes in CHD infants (including those from the total brain, cortex, white matter, frontal-, parietal- and temporal lobes and cerebellum) were not associated with one-year neurodevelopmental outcome when corrected for sex, postmenstrual age at time of MRI, MRI cohort and SES (see Supplementary Table S3 online). Similarly, no association was found between neonatal brain volume and one-year outcome in healthy controls (see Supplementary Table S4 online).

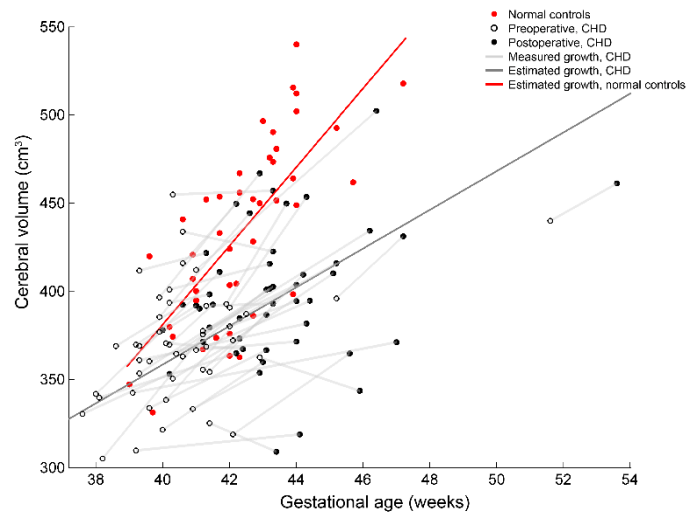

**Figure 2. Growth trajectories of total brain volumes of CHD infants compared to controls.** Red dots: brain volume of controls, Red line: cross-sectional brain growth of the controls. Black circles: preoperative brain volume of the CHD infants. Black dots filled: postoperative brain volume of the CHD infants. Black line: brain growth of CHD infants.

In contrast, *postoperative* regional brain volumes were associated with the one-year neurodevelopmental outcome. Specifically, the total brain, cortical, white matter, frontal lobe, temporal lobe and cerebellar volumes were positively associated with CCS. Furthermore, total brain, cortical, frontal lobe and temporal lobe volumes were associated with LCS (Table 4), when adjusted for sex, postmenstrual age at time of MRI, MRI cohort and SES. When additionally introducing length of ICU stay into the multiple regression model, CCS remained significantly associated with frontal (ß = 0.02, 95 % CI 0.00 to 0.04, p = 0.043), temporal (ß = 0.039, 95 % CI 0.01 to 0.07, p = 0.014) and cerebellar (ß = 0.056, 95 % CI 0.01 to 0.10, p = 0.018) volumes, whereas LCS remained independently predicted by cortical (ß = 0.0079, 95 % CI 0.00 to 0.02, p = 0.045), temporal (ß = 0.032, 95 % CI 0.00 to 0.06, p = 0.046) and frontal (ß = 0.021, 95 % CI 0.00 to 0.04, p = 0.027) brain volumes. In contrast, relative regional volumes (adjusted for total brain volume) did not correlate with neurodevelopmental outcome either in the pre- or in the postoperative MRI in CHD infants and in controls. In addition, neither total brain growth nor regional brain growth was predictive of any of the Bayley-III scores (see Supplementary Table S5 online).

*Table 4. Multiple linear regression models for each respective regional brain volume and neurodevelopmental outcome domain.* The model was adjusted for the covariates sex, postmenstrual age at MRI, MRI cohort and socioeconomic status. CCS, Bayley-III Cognitive Composite Score; LCS, Bayley-III Language Composite Score; TBV, total brain volume; CI, Confidence Interval. Only significant associations are shown. Further details on the model and non-significant results can be found in the Supplementary Table S6 online.

| Dependent variable | Independent variable* | ß Coefficient | 95 % CI | p |
|---|---|---|---|---|
| CCS | TBV | 0.0065 | 0.00 to 0.01 | 0.037 |
| CCS | Cortex | 0.0092 | 0.00 to 0.02 | 0.027 |
| CCS | WM | 0.012 | 0.00 to 0.02 | 0.036 |
| CCS | Frontal lobe | 0.022 | 0.00 to 0.04 | 0.027 |
| CCS | Temporal | 0.044 | 0.01 to 0.08 | 0.008 |
| CCS | Cerebellum | 0.06 | 0.01 to 0.11 | 0.017 |
| LCS | TBV | 0.0061 | 0.00 to 0.01 | 0.037 |
| LCS | Cortex | 0.0084 | 0.00 to 0.02 | 0.031 |
| LCS | Frontal lobe | 0.022 | 0.00 to 0.04 | 0.02 |
| LCS | Temporal lobe | 0.033 | 0.00 to 0.06 | 0.035 |

## Discussion

In this prospective cohort study, we examined the association between perioperative brain volumes and one-year neurodevelopmental outcome in children with severe CHD undergoing CPB. In our cohort, CHD infants had smaller global and regional brain volumes pre- and postoperatively. To the best of our knowledge, we are the first to show that postoperative total brain volume as well as cortical, frontal, temporal, white matter and cerebellar volume were predictive of neurodevelopmental outcome at one year, while, in contrast, preoperative brain volumes were not predictive of outcome. WMI was the predominant preoperative cerebral injury, whereas strokes and new postoperative lesions were rare. Importantly, both pre- and postoperative lesions were not predictive of neurodevelopmental outcome.

We found that brain volumes were smaller pre- and postoperatively in infants with CHD, and that all regional volumes were affected without a regional predilection. This is in line with the finding that the majority of relative regional brain volumes (after correcting for total brain volume) where similar between CHD infants and controls. These results are in accordance with those by von Rhein et al. [3] and Ortinau et al. [8], and most likely reflect a general developmental disturbance of brain growth and connectivity of foetal origin [31]. Studies on older children with CHD showed that brain volume reduction persists for all types of CHD from early childhood [32-34] until adolescence [35] and correlates with functional outcome. Importantly, we could not detect a predilection for a white or grey matter volume reduction. Watanabe et al. demonstrated a volume reduction that was most prominent in the grey matter, in particular in the frontal region, in children with CHD aged 13-15 months. In contrast, Rollins et al. found a reduction in white matter volume and a correlation with language function but not with cognitive or motor development at one year of age. However, in this study cerebral MRI was assessed at one year of age and alterations in brain growth may have become more apparent [33]. A study by Owen et al. demonstrated that the subcortical grey matter was reduced and the cerebrospinal fluid was increased in a mixed group of CHD infants and that these volume changes correlated with poor behavioural state regulation and poor visual orienting on the neonatal examination [4]. The difference in regional predilection may be due to differences in the distribution of studied cardiac diagnoses and medical variables such as postoperative course and complications.

We demonstrated that total postoperative brain volume and certain regional volumes were predictive of one-year cognitive and language outcome. Relative regional brain volumes, that is after controlling for total brain volume, however, did not correlate with outcome, as would be expected given the lack of a predilection for a certain brain region in the volume reduction.

While little is known about the association between neonatal brain volumes and outcome for the CHD populations, this association is well established for other at–risk populations. In children with perinatal asphyxia, hippocampal volume has been found to be reduced and associated with long-term outcome such as cognitive memory impairment as reviewed by de Haan et al. [36]. In preterm born infants, there is mounting evidence for a reduction in total and regional brain volumes with predictive value for neurodevelopmental outcome, which is particularly strong for the cerebellar volume [37-39]. However, in healthy infants in which no alteration or interruption of brain development occurs, the association between brain volume and outcome cannot be found.

We further examined potential risk factors for reduced pre- and postoperative brain volumes. While we found that neonatal characteristics such as head circumference at birth, gestational age and birth weight were related to pre- and postoperative total brain volume, longer length of postoperative ICU stay was related to lower postoperative brain volume. We assume that ICU stay is an indicator of disease complexity and is thus a strong factor influencing brain growth and by that neurodevelopmental outcome. However, in addition to this association, we found that, after adjusting for ICU stay, frontal, temporal and cerebellar brain volumes were independent predictors of cognitive outcome, and cortical frontal and temporal volumes of language outcome. These findings suggest that brain volume reduction is not just a mere mediator of the impact of ICU stay on outcome, but is also independently associated with outcome.

Our findings are consistent with those of Rollins et al., who examined CHD children at one year of age and found an association between a longer ICU stay and reduced grey matter volume as well as between lower intraoperative pCO2 and reduced total brain volume. No other risk factor for lower brain volume was identified in their study [33]. Interestingly, no other, and in particular no surgical or cardiac variable (i. e. extracorporeal circulation time, lowest temperature on bypass, aortic cross clamp time, univentricular CHD, dTGA vs. non-dTGA) was related to postoperative total brain volume. This observation is in line with the results from a large cohort of CHD infants in which measured intraoperative and postoperative factors accounted for less than 5% of the variances in one-year outcome while patient-specific and preoperative factors contributed much more to outcome [40].

Our study confirms that the predominant injuries in CHD infants are WMI and strokes, and new postoperative lesions are rare [2,11,41]. However, the prevalence of cerebral lesions may vary significantly depending on the type of CHD studied [9,18,41]. Consistent with the findings of Beca et al. [11], the cerebral lesions in our cohort were small, and therefore may lack an association with neurodevelopmental outcome, whereas others demonstrated a correlation between perioperative brain injuries and early childhood outcome [12,18,19]. This indicates that the presence or absence of an association between brain lesions and outcome depends on the distribution of CHD diagnoses, the rate of lesions and the method of quantification. While the presence or absence of lesions may not be useful in predicting outcome, the volume of white matter injury may better reflect the burden of injury and by that may be predictive of outcome as has been recently demonstrated by Peyvandi et al. [12].

Neurodevelopmental outcome in the CHD population was lower in all domains but within the normal range. Our outcomes are in line with the findings of Andropoulos et al. who also used the Bayley-III in their cohort [18], although earlier. However, the test version needs to be considered when comparing outcome results among studies. It has been shown that the Bayley-III generates significantly higher scores compared to the Bayley-II, and thus overestimates functional outcome [42].

Brain growth between pre- and postoperative scans was reduced in CHD infants, but not predictive of outcome. Brain volume appears to increase in healthy infants at a higher rate than in CHD patients in their first weeks of life. This observation may be linked to the persisting developmental impairment of the central nervous system even after corrective surgery. We hypothesize that a stronger association between early brain growth rate and outcome would be found if a longer period was examined, as has been shown in the population of preterm born infants by Cheong et al. [43]. They could demonstrate that a poor postnatal head growth, as a surrogate marker for brain growth, became evident between birth and two years of age and was predictive of neurodevelopmental outcome [43].

We suggest that a multiple-hit-theory could explain why the preoperatively measured brain volumes did not correlate with outcome, whereas the postoperative volumes did. In this model, the "first hit" takes place during prenatal life, caused by different factors such as hypoperfusion and neuronal disturbances, while the second hit occurs during the perioperative phase. The additive or potentiating effect of these events may only become evident during later development, and the inter-individual variability of brain growth trajectories will become larger in later life. A more precise assessment of a possible divergence in such growth trajectories would necessitate longitudinal neurodevelopmental evaluations and repeated MR imaging with volumetry over multiple months after surgery.

This study has several strengths. Pre- and postoperative MRIs could be obtained in a large number of patients and brain volumes were measured with an automated pipeline in a fast and reliable fashion. The follow up rate of one-year neurodevelopmental follow up was excellent and assessment was performed with a detailed, standardized developmental test (Bayley-III). However, there are also considerable limitations. It is important to note that the predictive value of the Bayley-III assessment for long-term neurodevelopmental outcome is limited [44]. Later neurodevelopmental follow up is warranted to confirm our results. Subgroup analyses or stratification of the linear regression results by the CHD diagnoses would yield more specific results, but power to detect differences is limited due to sample size. Moreover, our study population is a heterogeneous, clinically diverse sample composed of patients with various CHD diagnoses. Furthermore, we performed multiple regression analyses to investigate the predictive value of regional brain volumes, leading to multiple hypothesis tests which must be considered when interpreting the results as we did not correct for multiple testing. A technical limitation of our study is that the MRI scanner was upgraded during the study. While the same pulse sequence was used, the upgrade could theoretically affect the MRI contrast, potentially introducing a bias into the brain volumes calculated using the automated approach. To control for the confounding effect of MRI scanner software type, we corrected for this variable in all statistical analyses, and we additionally validated the volumes derived using the automated approach with those from a manual gold-standard method. However, the in-house developed automatic work-flow should also be compared to alternative automated approaches in the future [29]. During the validation of the anatomical prior based automatic segmentation approach (Supplementary Note), we found high variability for total cortical volume, which limits the generalisability of our volumetric findings. More reliable and reproducible results can be expected for lobar white matter, total brain volume and cerebellar volume estimates, which was confirmed by higher correlation and spatial overlap of these volume estimates with a more recently published automated approach (dHCP pipeline, more details are found in Supplementary Note), and higher correlation to manual stereology.

In sum, we found that postoperative regional brain volumes of children with CHD were smaller than those of controls and were predictive of neurodevelopmental outcome while preoperative brain volumes were not. Our findings add to the evidence of an association between neonatal brain volumes and neurodevelopmental outcome across at-risk populations. Our results underscore the importance of early neuroimaging assessments in infants with severe CHD, providing measures such as the total brain volume which could serve as a biomarker for altered brain development in combination with a more complicated postoperative course.

## Acknowledgements

We thank Ulrike Held for her statistical support. Financial support for A.J. was provided by the OPO Foundation, the Foundation for Research in Science and the Humanities at the University of Zurich, the Hasler Foundation and the Forschungszentrum für das Kind Grant (FZK). A.J. and R.T. are supported by the EMDO Foundation, grant number 928. E.M. and M.F. are supported by the Mäxi Foundation.

## Research group Heart and Brain

Rabia Liamlahi, Annette Hackenberg, Oliver Kretschmar, Christian Kellenberger, Christoph Bürki, Markus Weiss

## This manuscript is the author's original version.

**Supplementary Information**

**Postoperative brain volumes are associated with one-year neurodevelopmental outcome in children with severe congenital heart disease**

Eliane Meuwly, Maria Feldmann, Walter Knirsch, Michael von Rhein, Kelly Payette, Hitendu Dave, Ruth O'Gorman Tuura, Raimund Kottke, Cornelia Hagmann, Research Group Heart and Brain, Beatrice Latal and András Jakab

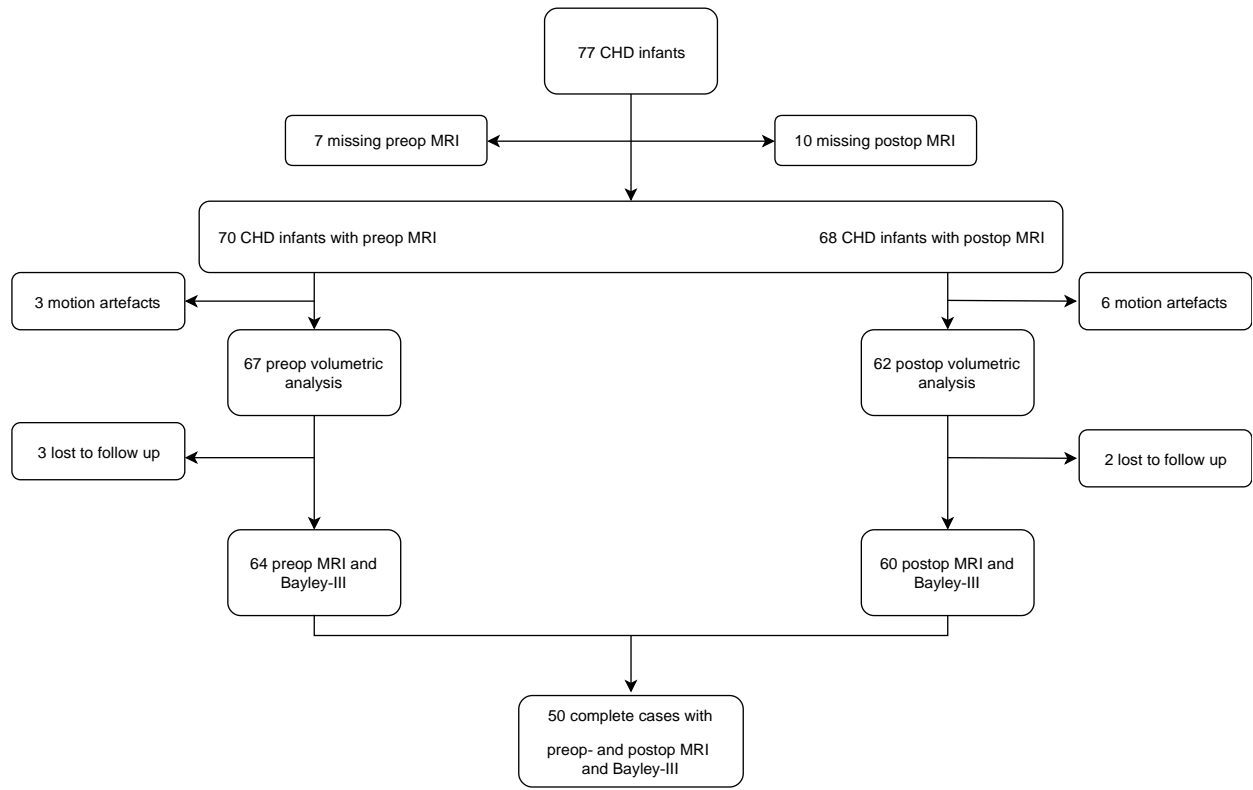

**Supplementary Figure S1**. **Flowchart of enrolled CHD infants.** CHD, congenital heart disease; postop, postoperative; preop, preoperative; MRI, magnetic resonance imaging, Bayley-III, Bayley Scales of Infant and Toddler Development, Third Edition.

**Supplementary Table S1. Correlation between perinatal and intraoperative risk factors and postoperative total brain volume.** Pearson correlation of risk factors and postoperative total brain volume. $r$, correlation coefficient; CI, confidence interval; ECC, extracorporeal circulation time; ACC, aortic cross clamp time.

| Risk factor | $r$ | 95 % CI | $p$ |
|---|---|---|---|
| Apgar 5 min | -0.017 | -0.26; 0.23 | 0.90 |
| ECC | 0.072 | -0.19; 0.32 | 0.58 |
| Lowest temperature on bypass | 0.14 | -0.13; 0.38 | 0.32 |
| ACC | 0.079 | -0.18; 0.32 | 0.55 |

**Supplementary Table S2. Comparison of pre- and postoperative brain volumes in CHD infants.** Paired t test. CI, confidence interval.

| Brain volume | Mean difference (cm$^3$) | 95 % CI | $p$ |
|---|---|---|---|
| Total brain volume | 30.49 | 18.42; 42.56 | < 0.0001 |
| Cortex | 25.97 | 15.91; 36.03 | < 0.0001 |
| White matter | 10.45 | 4.29; 16.60 | 0.0013 |
| Frontal lobe | 7.67 | 4.36; 10.98 | < 0.0001 |
| Parietal lobe | 4.55 | 2.16; 6.94 | 0.0003 |
| Occipital lobe | 3.61 | 1.87; 5.35 | 0.0001 |
| Temporal lobe | 5.10 | 3.28; 6.93 | < 0.0001 |
| Cerebellum | 3.94 | 2.74; 5.14 | < 0.0001 |

**Supplementary Table S3. Association between preoperative neonatal brain volume and one-year neurodevelopmental outcome in CHD infants.** Multiple linear regression models for each respective preoperative brain volume and neurodevelopmental outcome domain in CHD infants. ß, ß coefficient of the covariates; CI, confidence interval; Adj, adjusted; TBV, total brain volume; PMA, postmenstrual age; MRI, magnetic resonance imaging; SES, socioeconomic status. p Model corresponds to p-value of the linear regression model.

| | Cognitive Composite Score | | | Language Composite Score | | | Motor Composite Score | | |
|---|---|---|---|---|---|---|---|---|---|
| | ß | 95 % CI | p | ß | 95 % CI | p | ß | 95 % CI | p |
| Intercept | 9,52 | 5.60 ; 13.44 | < 0.0001 | 8,34 | 4.61 ; 12.06 | < 0.0001 | 7,3 | 2.99 ; 11.62 | 0.001 |
| **TBV** | 0,0038 | -0.00 ; 0.01 | 0,13 | 0,0032 | -0.00 ; 0.01 | 0,17 | 0,00068 | -0.00 ; 0.01 | 0,8 |
| Sex | -0,33 | -0.77 ; 0.11 | 0,14 | -0,23 | -0.65 ; 0.18 | 0,27 | -0,11 | -0.59 ; 0.38 | 0,66 |
| PMA at MRI | -0,028 | -0.12 ; 0.06 | 0,53 | -0,016 | -0.10 ; 0.07 | 0,71 | 0,031 | -0.07 ; 0.13 | 0,53 |
| MRI Cohort | 0,2 | -0.18 ; 0.58 | 0,29 | 0,24 | -0.12 ; 0.60 | 0,19 | 0,41 | -0.01 ; 0.83 | 0,055 |
| SES | 0,091 | 0.00 ; 0.18 | 0,045 | 0,1 | 0.02 ; 0.19 | 0,02 | 0,068 | -0.03 ; 0.17 | 0,17 |
| | Adj R²: 0.073 | | | Adj R²: 0.091 | | | Adj R²: 0.28 | | |
| | p Model: 0.12 | | | p Model: 0.081 | | | p Model: 0.28 | | |
| Intercept | 10,1 | 6.27 ; 13.94 | < 0.0001 | 8,84 | 5.18 ; 12.49 | < 0.0001 | 7,43 | 3.15 ; 11.72 | 0.001 |
| **Cortex** | 0,0063 | 0.00 ; 0.01 | 0.047 | 0,0054 | -0.00 ; 0.01 | 0.073 | 0,0018 | -0.01 ; 0.01 | 0,61 |
| Sex | -0,38 | -0.81 ; 0.06 | 0.089 | -0,27 | -0.69 ; 0.14 | 0,19 | -0,13 | -0.62 ; 0.36 | 0,6 |
| PMA at MRI | -0,044 | -0.14 ; 0.05 | 0,34 | -0,03 | -0.12 ; 0.06 | 0,5 | 0,025 | -0.08 ; 0.13 | 0,63 |
| MRI Cohort | 0,1 | -0.29 ; 0.49 | 0,6 | 0,1 | -0.22 ; 0.53 | 0,41 | 0,38 | -0.06 ; 0.81 | 0,088 |
| SES | 0,092 | 0.00 ; 0.18 | 0,039 | 0,1 | 0.02 ; 0.18 | 0,017 | 0,068 | -0.03 ; 0.16 | 0,17 |
| | Adj R²: 0.10 | | | Adj R²: 0.12 | | | Adj R²: 0.031 | | |
| | p Model: 0.061 | | | p Model: 0.047 | | | p Model: 0.26 | | |
| Intercept | 9,18 | 5.12 ; 13.25 | < 0.0001 | 8,05 | 4.19 ; 11.91 | 0,0001 | 7,45 | 3.01 ; 11.89 | 0.001 |
| **White matter** | 0,006 | -0.00 ; 0.02 | 0,21 | 0,0051 | -0.00 ; 0.01 | 0,25 | -0,00072 | -0.01 ; 0.01 | 0,89 |
| Sex | -0,3 | -0.74 ; 0.13 | 0,17 | -0,21 | -0.62 ; 0.20 | 0,31 | -0,082 | -0.56 ; 0.40 | 0,73 |
| PMA at MRI | -0,016 | -0.10 ; 0.07 | 0,71 | -0,0055 | -0.09 ; 0.08 | 0,9 | 0,036 | -0.06 ; 0.13 | 0,45 |
| MRI Cohort | 0,29 | -0.11 ; 0.68 | 0,15 | 0,31 | -0.06 ; 0.69 | 0,1 | 0,4 | -0.03 ; 0.84 | 0,065 |
| SES | 0,091 | 0.00 ; 0.18 | 0,046 | 0,1 | 0.02 ; 0.19 | 0,021 | 0,07 | -0.03 ; 0.17 | 0,16 |
| | Adj R²: 0.061 | | | Adj R²: 0.080 | | | Adj R²: 0.026 | | |
| | p Model: 0.15 | | | p Model: 0.10 | | | p Model: 0.28 | | |
| Intercept | 9,49 | 5.56 ; 13.42 | < 0.0001 | 8,24 | 4.54 ; 11.94 | < 0.0001 | 7,44 | 3.12 ; 11.76 | 0.001 |
| **Frontal lobe** | 0,014 | -0.00 ; 0.03 | 0,14 | 0,014 | -0.00 ; 0.03 | 0,1 | -0,0028 | -0.02 ; 0.02 | 0,79 |
| Sex | -0,34 | -0.79 ; 0.10 | 0,13 | -0,26 | -0.68 ; 0.16 | 0,22 | -0,069 | -0.56 ; 0.42 | 0,78 |
| PMA at MRI | -0,029 | -0.12 ; 0.06 | 0,52 | -0,021 | -0.11 ; 0.06 | 0,63 | 0,04 | -0.06 ; 0.14 | 0,43 |
| MRI Cohort | 0,25 | -0.13 ; 0.63 | 0,19 | 0,29 | -0.07 ; 0.65 | 0,11 | 0,41 | -0.01 ; 0.83 | 0,056 |
| SES | 0,089 | -0.00 ; 0.18 | 0,051 | 0,097 | 0.01 ; 0.18 | 0,024 | 0,071 | -0.03 ; 0.17 | 0,15 |
| | Adj R²: 0.071 | | | Adj R²: 0.10 | | | Adj R²: 0.027 | | |
| | p Model: 0.12 | | | p Model: 0.060 | | | p Model: 0.27 | | |
| Intercept | 9,3 | 5.43 ; 13.17 | < 0.0001 | 8,39 | 4.61 ; 12.17 | < 0.0001 | 7,15 | 2.84 ; 11.46 | 0.002 |
| **Parietal lobe** | 0,022 | 0.00 ; 0.04 | 0,046 | 0,0095 | -0.01 ; 0.03 | 0,38 | 0,0086 | -0.02 ; 0.03 | 0,48 |
| Sex | -0,33 | -0.75 ; 0.09 | 0,12 | -0,19 | -0.61 ; 0.22 | 0,35 | -0,12 | -0.60 ; 0.35 | 0,6 |
| PMA at MRI | -0,028 | -0.12 ; 0.06 | 0,52 | -0,007 | -0.09 ; 0.08 | 0,87 | 0,027 | -0.07 ; 0.12 | 0,57 |
| MRI Cohort | 0,29 | -0.09 ; 0.67 | 0,13 | 0,29 | -0.08 ; 0.65 | 0,13 | 0,44 | 0.02 ; 0.86 | 0,042 |
| SES | 0,093 | 0.01 ; 0.18 | 0,04 | 0,1 | 0.02 ; 0.19 | 0,018 | 0,068 | -0.03 ; 0.16 | 0,17 |
| | Adj R²: 0.10 | | | Adj R²: 0.071 | | | Adj R²: 0.036 | | |
| | p Model: 0.06 | | | p Model: 0.12 | | | p Model: 0.24 | | |
| Intercept | 9,78 | 5.78 ; 13.77 | < 0.0001 | 8,48 | 4.72 ; 12.25 | < 0.0001 | 7,37 | 3.07 ; 11.68 | 0.001 |
| **Occipital lobe** | 0,0078 | -0.03 ; 0.04 | 0,66 | 0,014 | -0.02 ; 0.05 | 0,41 | -0,00057 | -0.04 ; 0.04 | 0,98 |
| Sex | -0,26 | -0.70 ; 0.18 | 0,24 | -0,18 | -0.60 ; 0.23 | 0,37 | -0,088 | -0.56 ; 0.38 | 0,71 |
| PMA at MRI | -0,011 | -0.10 ; 0.08 | 0,81 | -0,0038 | -0.09 ; 0.08 | 0,93 | 0,035 | -0.06 ; 0.13 | 0,46 |
| MRI Cohort | 0,21 | -0.18 ; 0.60 | 0,28 | 0,24 | -0.13 ; 0.60 | 0,2 | 0,41 | -0.01 ; 0.83 | 0,054 |
| SES | 0,095 | 0.00 ; 0.19 | 0,039 | 0,1 | 0.02 ; 0.19 | 0,018 | 0,069 | -0.03 ; 0.17 | 0,16 |
| | Adj R²: 0.034 | | | Adj R²: 0.069 | | | Adj R²: 0.026 | | |
| | p Model: 0.25 | | | p Model: 0.13 | | | p Model: 0.28 | | |
| Intercept | 9,32 | 5.42 ; 13.21 | < 0.0001 | 8,14 | 4.44 ; 11.84 | < 0.0001 | 7,34 | 3.01 ; 11.68 | 0.001 |
| **Temporal lobe** | 0,027 | -0.00 ; 0.06 | 0,068 | 0,024 | -0.00 ; 0.05 | 0,082 | 0,0012 | -0.03 ; 0.03 | 0,94 |
| Sex | -0,37 | -0.81 ; 0.07 | 0,098 | -0,27 | -0.69 ; 0.15 | 0,2 | -0,096 | -0.59 ; 0.39 | 0,7 |
| PMA at MRI | -0,029 | -0.12 ; 0.06 | 0,52 | -0,017 | -0.10 ; 0.07 | 0,68 | 0,034 | -0.06 ; 0.13 | 0,48 |
| MRI Cohort | 0,22 | -0.15 ; 0.60 | 0,24 | 0,26 | -0.10 ; 0.61 | 0,15 | 0,41 | -0.01 ; 0.83 | 0,052 |
| SES | 0,087 | -0.00 ; 0.18 | 0,052 | 0,097 | 0.01 ; 0.18 | 0,024 | 0,069 | -0.03 ; 0.17 | 0,16 |
| | Adj R²: 0.093 | | | Adj R²: 0.11 | | | Adj R²: 0.28 | | |
| | p Model: 0.077 | | | p Model: 0.051 | | | p Model: 0.28 | | |
| Intercept | 9,91 | 6.03 ; 13.79 | < 0.0001 | 8,65 | 4.90 ; 12.40 | < 0.0001 | 7,39 | 3.14 ; 11.65 | 0.001 |
| **Cerebellum** | 0,04 | -0.01 ; 0.09 | 0,099 | 0,017 | -0.03 ; 0.06 | 0,47 | 0,022 | -0.03 ; 0.07 | 0,4 |
| Sex | -0,29 | -0.71 ; 0.14 | 0,18 | -0,17 | -0.58 ; 0.24 | 0,4 | -0,11 | -0.58 ; 0.35 | 0,62 |
| PMA at MRI | -0,037 | -0.13 ; 0.06 | 0,42 | -0,01 | -0.10 ; 0.08 | 0,82 | 0,019 | -0.08 ; 0.12 | 0,71 |
| MRI Cohort | 0,16 | -0.22 ; 0.55 | 0,4 | 0,23 | -0.14 ; 0.60 | 0,22 | 0,38 | -0.04 ; 0.80 | 0,078 |
| SES | 0,11 | 0.02 ; 0.20 | 0,019 | 0,11 | 0.02 ; 0.20 | 0,013 | 0,075 | -0.02 ; 0.17 | 0,13 |
| | Adj R²: 0.082 | | | Adj R²: 0.066 | | | Adj R²: 0.04 | | |
| | p Model: 0.097 | | | p Model: 0.14 | | | p Model: 0.22 | | |

**Supplementary Table S4. Association between neonatal brain volume and one-year neurodevelopmental outcome in healthy controls**. Multiple linear regression models for total brain volume and each neurodevelopmental outcome domain in healthy controls. ß, ß coefficient of the covariates; CI, confidence interval; Adj, adjusted; TBV, total brain volume; PMA, postmenstrual age; MRI, magnetic resonance imaging; SES, socioeconomic status. p Model corresponds to p-value of the linear regression model.

| | Cognitive Composite Score | | | Language Composite Score | | | Motor Composite Score | | |
|---|---|---|---|---|---|---|---|---|---|
| | ß | 95 % CI | p | ß | 95 % CI | p | ß | 95 % CI | p |
| Intercept | 8,47 | 1.68; 15.26 | 0.016 | 8,35 | 1.50; 15.20 | 0.019 | 9 | 1.65; 16.36 | 0.018 |
| **TBV** | 0,0045 | -0.00 ; 0.01 | 0,2 | -0,00084 | -0.01 ; 0.01 | 0,81 | 0,0014 | -0.01 ; 0.01 | 0,71 |
| Sex | -0,26 | -0.70 ; 0.18 | 0,24 | 0,23 | -0.22 ; 0.67 | 0,3 | -0.053 | -0.53 ; 0.43 | 0,82 |
| PMA at MRI | 0,0077 | -0.21 ; 0.22 | 0,94 | 0.051 | -0.17 ; 0.27 | 0,63 | -0,0047 | -0.24 ; 0.23 | 0,97 |
| MRI Cohort | -0,26 | -0.67 ; 0.15 | 0,21 | 0,23 | -0.18 ; 0.65 | 0,25 | 0,2 | -0.24 ; 0.64 | 0,36 |
| SES | 0.029 | -0.11 ; 0.17 | 0,67 | -0.048 | -0.19 ; 0.09 | 0,48 | 0.062 | -0.09 ; 0.21 | 0,4 |
| | Adj R²: 0.096 | | | Adj R²: -0.077 | | | Adj R²: -0.091 | | |
| | p Model: 0.18 | | | p Model: 0.73 | | | p Model: 0.79 | | |

**Supplementary Table S5. Association between regional brain growth and one-year neurodevelopmental outcome in CHD infants.** Multiple linear regression models for total and regional brain growth in cm³/week and each neurodevelopmental outcome domain in CHD infants. ß, ß coefficient of the covariates; CI, confidence interval; Adj, adjusted; TBV, total brain volume; MRI, magnetic resonance imaging; SES, socioeconomic status. p Model corresponds to p-value of the linear regression model.

| | Cognitive Composite Score | | | Language Composite Score | | | Motor Composite Score | | |
|---|---|---|---|---|---|---|---|---|---|
| | ß | 95 % CI | p | ß | 95 % CI | p | ß | 95 % CI | p |
| Intercept | 9.69 | 8.75 ; 10.63 | < 0.0001 | 8,59 | 7.66 ; 9.52 | < 0.0001 | 8,86 | 7.89 ; 9.84 | < 0.0001 |
| **Total brain growth** | -0.001 | -0.01 ; 0.01 | 0.85 | -0.005 | -0.02 ; 0.01 | 0,37 | 0.001 | -0.01 ; 0.01 | 0,86 |
| Sex | -0.14 | -0.62 ; 0.33 | 0.55 | -0,036 | -0.50 ; 0.43 | 0,88 | 0,00043 | -0.49 ; 0.49 | 1 |
| MRI Cohort | 0.23 | -0.18 ; 0.63 | 0.27 | 0,29 | -0.12 ; 0.69 | 0,16 | 0,39 | -0.04 ; 0.81 | 0,073 |
| SES | 0.079 | -0.02 ; 0.17 | 0.098 | 0,11 | 0.02 ; 0.21 | 0,017 | 0.065 | -0.03 ; 0.16 | 0,19 |
| | | Adj R²: 0.0038 | | | Adj R²: 0.084 | | | Adj R²: 0.018 | |
| | | *p* Model: 0.39 | | | *p* Model: 0.092 | | | *p* Model: 0.31 | |
| Intercept | 9,69 | 8.75 ; 10.64 | < 0.0001 | 8,61 | 7.69 ; 9.53 | < 0.0001 | 8,86 | 7.88 ; 9.84 | < 0.0001 |
| **Cortical growth** | -0,0029 | -0.02 ; 0.01 | 0,67 | 0,0076 | -0.02 ; 0.01 | 0,26 | 0,0017 | -0.01 ; 0.02 | 0,81 |
| Sex | -0,13 | -0.60 ; 0.33 | 0,57 | -0,036 | -0.49 ; 0.42 | 0,88 | 0,00049 | -0.49 ; 0.48 | 1 |
| MRI Cohort | 0,22 | -0.19 ; 0.63 | 0,28 | 0,28 | -0.12 ; 0.68 | 0,17 | 0,39 | -0.04 ; 0.81 | 0,072 |
| SES | 0,079 | -0.01 ; 0.17 | 0.096 | 0,11 | 0.02 ; 0.21 | 0,017 | 0.065 | -0.03 ; 0.16 | 0,19 |
| | | Adj R²: 0.0071 | | | Adj R²: 0.094 | | | Adj R²: 0.019 | |
| | | *p* Model: 0.37 | | | *p* Model: 0.076 | | | *p* Model: 0.31 | |
| Intercept | 9,69 | 8.74 ; 10.64 | < 0.0001 | 8,56 | 7.62 ; 9.50 | < 0.0001 | 8,86 | 7.87 ; 9.85 | < 0.0001 |
| **White matter growth** | 0,0007 | -0.02 ; 0.02 | 0,95 | 0,0071 | -0.03 ; 0.01 | 0,49 | -0,0004 | -0.02 ; 0.02 | 0,97 |
| Sex | -0,16 | -0.64 ; 0.31 | 0,5 | -0,051 | -0.52 ; 0.42 | 0,83 | 0,016 | -0.48 ; 0.51 | 0,95 |
| MRI Cohort | 0,23 | -0.18 ; 0.64 | 0,27 | 0,28 | -0.13 ; 0.69 | 0,17 | 0,38 | -0.04 ; 0.81 | 0,077 |
| SES | 0,078 | -0.02 ; 0.17 | 0,11 | 0,12 | 0.02 ; 0.21 | 0,016 | 0,066 | -0.03 ; 0.16 | 0,19 |
| | | Adj R²: 0.0031 | | | Adj R²: 0.078 | | | Adj R²: 0.018 | |
| | | *p* Model: 0.40 | | | *p* Model: 0.11 | | | *p* Model: 0.31 | |
| Intercept | 9,69 | 8.75 ; 10.63 | < 0.0001 | 8,59 | 7.66 ; 9.53 | < 0.0001 | 8,86 | 7.88 ; 9.84 | < 0.0001 |
| **Frontal lobe growth** | 0,00034 | -0.04 ; 0.04 | 0,99 | -0,012 | -0.05 ; 0.03 | 0,56 | 0,0017 | -0.04 ; 0.05 | 0,94 |
| Sex | -0,16 | -0.62 ; 0.31 | 0,5 | -0,067 | -0.53 ; 0.40 | 0,77 | 0,009 | -0.48 ; 0.49 | 0,97 |
| MRI Cohort | 0,23 | -0.18 ; 0.63 | 0,27 | 0,29 | -0.11 ; 0.70 | 0,15 | 0,38 | -0.04 ; 0.81 | 0,074 |
| SES | 0,078 | -0.02 ; 0.17 | 0,1 | 0,11 | 0.02 ; 0.21 | 0,018 | 0,065 | -0.03 ; 0.16 | 0,19 |
| | | Adj R²: 0.0030 | | | Adj R²: 0.075 | | | Adj R²: 0.018 | |
| | | *p* Model: 0.40 | | | *p* Model: 0.11 | | | *p* Model: 0.31 | |
| Intercept | 9,67 | 8.72 ; 10.62 | < 0.0001 | 8,54 | 7.61 ; 9.48 | < 0.0001 | 8,85 | 7.86 ; 9.83 | < 0.0001 |
| **Parietal lobe growth** | -0,0081 | -0.06 ; 0.05 | 0,76 | -0,025 | -0.08 ; 0.03 | 0,34 | -0,0074 | -0.06 ; 0.05 | 0,79 |
| Sex | -0,13 | -0.61 ; 0.35 | 0,58 | -0,023 | -0.50 ; 0.45 | 0,92 | 0,036 | -0.46 ; 0.54 | 0,89 |
| MRI Cohort | 0,22 | -0.20 ; 0.63 | 0,3 | 0,26 | -0.14 ; 0.67 | 0,2 | 0,37 | -0.05 ; 0.80 | 0,086 |
| SES | 0,08 | -0.01 ; 0.18 | 0.094 | 0,12 | 0.02 ; 0.21 | 0,014 | 0,067 | -0.03 ; 0.17 | 0,18 |
| | | Adj R²: 0.0050 | | | Adj R²: 0.087 | | | Adj R²: 0.020 | |
| | | *p* Model: 0.39 | | | *p* Model: 0.088 | | | *p* Model: 0.31 | |
| Intercept | 9,7 | 8.76 ; 10.64 | < 0.0001 | 8,59 | 7.65 ; 9.52 | < 0.0001 | 8,87 | 7.89 ; 9.85 | < 0.0001 |
| **Occipital lobe growth** | 0,016 | -0.06 ; 0.09 | 0,66 | -0,017 | -0.09 ; 0.05 | 0,64 | 0,013 | -0.06 ; 0.09 | 0,73 |
| Sex | -0,18 | -0.64 ; 0.28 | 0,44 | -0,075 | -0.54 ; 0.39 | 0,75 | -0,0054 | -0.49 ; 0.48 | 0,98 |
| MRI Cohort | 0,24 | -0.17 ; 0.65 | 0,25 | 0,28 | -0.12 ; 0.69 | 0,17 | 0,39 | -0.03 ; 0.82 | 0,07 |
| SES | 0,076 | -0.02 ; 0.17 | 0,11 | 0,11 | 0.02 ; 0.21 | 0,018 | 0,064 | -0.03 ; 0.16 | 0,2 |
| | | Adj R²: 0.0074 | | | Adj R²: 0.073 | | | Adj R²: 0.021 | |
| | | *p* Model: 0.37 | | | *p* Model: 0.12 | | | *p* Model: 0.30 | |
| Intercept | 9,68 | 8.74 ; 10.62 | < 0.0001 | 8,61 | 7.69 ; 9.54 | < 0.0001 | 8,85 | 7.88 ; 9.83 | < 0.0001 |
| **Temporal lobe growth** | 0,015 | -0.06 ; 0.09 | 0,67 | -0,035 | -0.11 ; 0.04 | 0,32 | 0,023 | -0.05 ; 0.10 | 0,53 |
| Sex | -0,19 | -0.66 ; 0.29 | 0,43 | -0,031 | -0.50 ; 0.43 | 0,89 | -0,031 | -0.52 ; 0.46 | 0,9 |
| MRI Cohort | 0,23 | -0.17 ; 0.64 | 0,25 | 0,28 | -0.12 ; 0.68 | 0,17 | 0,4 | -0.03 ; 0.82 | 0,066 |
| SES | 0,078 | -0.02 ; 0.17 | 0,1 | 0,11 | 0.02 ; 0.21 | 0,017 | 0,064 | -0.03 ; 0.16 | 0,19 |
| | | Adj R²: 0.0070 | | | Adj R²: 0.089 | | | Adj R²: 0.027 | |
| | | *p* Model: 0.37 | | | *p* Model: 0.085 | | | *p* Model: 0.27 | |
| Intercept | 9,71 | 8.76 ; 10.65 | < 0.0001 | 8,58 | 7.65 ; 9.52 | < 0.0001 | 8,87 | 7.89 ; 9.85 | < 0.0001 |
| **Cerebellar growth** | 0,033 | -0.09 ; 0.16 | 0,59 | -0,027 | -0.15 ; 0.10 | 0,67 | 0,0079 | -0.12 ; 0.14 | 0,9 |
| Sex | -0,2 | -0.69 ; 0.28 | 0,4 | -0,061 | -0.54 ; 0.42 | 0,8 | 0,0024 | -0.50 ; 0.51 | 0,99 |

| | Cognitive Composite Score | | | Language Composite Score | | | Motor Composite Score | | |
|---|---|---|---|---|---|---|---|---|---|
| MRI Cohort | 0,24 | -0.17 ; 0.65 | 0,25 | 0,29 | -0.12 ; 0.69 | 0.16 | 0,39 | -0.04 ; 0.81 | 0.074 |
| SES | 0.075 | -0.02 ; 0.17 | 0.12 | 0,11 | 0.02 ; 0.21 | 0.018 | 0.064 | -0.03 ; 0.16 | 0,19 |
| | Adj R²: 0.0094 | | | Adj R²: 0.072 | | | Adj R²: 0.018 | | |
| | p Model: 0.36 | | | p Model: 0.12 | | | p Model: 0.31 | | |

**Supplementary Table S6. Association between postoperative neonatal brain volume and one-year neurodevelopmental outcome in CHD infants.** Multiple linear regression models for each respective postoperative regional brain volume and neurodevelopmental outcome domain in CHD infants. ß, ß coefficient of the covariates; CI, confidence interval; Adj, adjusted; TBV, total brain volume; PMA, postmenstrual age; MRI, magnetic resonance imaging; SES, socioeconomic status. p Model corresponds to p-value of the linear regression model. P values of brain volumes independently associated with outcome are printed in bold.

| | Cognitive Composite Score | | | Language Composite Score | | | Motor Composite Score | | |
|---|---|---|---|---|---|---|---|---|---|
| | ß | 95% CI | p | ß | 95% CI | p | ß | 95% CI | p |
| Intercept | 7,33 | 3.35 ; 11.31 | 0,0005 | 7,62 | 3.91 ; 11.32 | 0,0001 | 3,94 | -0.02 ; 7.90 | 0.051 |
| **TBV** | 0,0065 | 0.00 ; 0.01 | **0.037** | 0,0061 | 0.00 ; 0.01 | **0.037** | 0,0048 | -0.00 ; 0.01 | 0,12 |
| Sex | -0,33 | -0.85 ; 0.18 | 0,2 | -0,34 | -0.82 ; 0.15 | 0,17 | -0,16 | -0.67 ; 0.34 | 0,54 |
| PMA at MRI | -0,0046 | -0.11 ; 0.10 | 0,93 | -0,03 | -0.13 ; 0.07 | 0,54 | 0,069 | -0.03 ; 0.17 | 0,18 |
| MRI Cohort | -0,022 | -0.42 ; 0.38 | 0,91 | 0,2 | -0.18 ; 0.57 | 0,3 | 0,26 | -0.14 ; 0.65 | 0,2 |
| SES | 0.095 | 0.00 ; 0.19 | 0.049 | 0,12 | 0.03 ; 0.21 | 0.013 | 0.085 | -0.01 ; 0.18 | 0.076 |
| | Adj. R²: 0.11 | | | Adj R²: 0.15 | | | Adj R²: 0.15 | | |
| | p Model: 0.049 | | | p Model: 0.017 | | | p Model: 0.018 | | |
| Intercept | 8,47 | 4.34 ; 12.61 | 0,0001 | 8,66 | 4.80 ; 12.52 | < 0.0001 | 4,86 | 0.74 ; 8.97 | 0.022 |
| **Cortex** | 0,0092 | 0.00 ; 0.02 | **0.027** | 0,0084 | 0.00 ; 0.02 | **0.031** | 0,0073 | -0.00 ; 0.02 | 0.075 |
| Sex | -0,34 | -0.84 ; 0.17 | 0,19 | -0,33 | -0.81 ; 0.14 | 0,17 | -0,17 | -0.67 ; 0.33 | 0,5 |
| PMA at MRI | -0,028 | -0.14 ; 0.08 | 0,62 | -0,05 | -0.16 ; 0.06 | 0,34 | 0,047 | -0.06 ; 0.16 | 0,4 |
| MRI Cohort | -0,14 | -0.57 ; 0.28 | 0,5 | 0,085 | -0.32 ; 0.49 | 0,67 | 0,16 | -0.26 ; 0.58 | 0,46 |
| SES | 0.095 | -0.00 ; 0.19 | 0.049 | 0,12 | 0.03 ; 0.21 | 0.011 | 0.084 | -0.01 ; 0.18 | 0.078 |
| | Adj R²: 0.12 | | | Adj R²: 0.16 | | | Adj R²: 0.16 | | |
| | p Model: 0.039 | | | p Model: 0.015 | | | p Model: 0.013 | | |
| Intercept | 5,79 | 1.64 ; 9.94 | 0.007 | 6,26 | 2.37 ; 10.16 | 0.002 | 2,77 | -1.36 ; 6.90 | 0,18 |
| **White matter** | 0.012 | 0.00 ; 0.02 | **0.036** | 0,01 | -0.00 ; 0.02 | 0.057 | 0,0093 | -0.00 ; 0.02 | 0,11 |
| Sex | -0,31 | -0.81 ; 0.19 | 0,23 | -0,3 | -0.78 ; 0.18 | 0,21 | -0,14 | -0.63 ; 0.36 | 0,58 |
| PMA at MRI | 0.033 | -0.06 ; 0.12 | 0,46 | 0,0082 | -0.08 ; 0.09 | 0,85 | 0,097 | 0.01 ; 0.19 | 0.035 |
| MRI Cohort | 0,18 | -0.24 ; 0.60 | 0,39 | 0,38 | -0.02 ; 0.78 | 0,06 | 0,41 | -0.01 ; 0.83 | 0.055 |
| SES | 0.089 | -0.01 ; 0.18 | 0.068 | 0,11 | 0.02 ; 0.21 | 0.019 | 0,08 | -0.02 ; 0.18 | 0.097 |
| | Adj R²: 0.11 | | | Adj R²: 0.14 | | | Adj R²: 0.15 | | |
| | p Model: 0.047 | | | p Model: 0.023 | | | p Model: 0.017 | | |
| Intercept | 6,25 | 2.23 ; 10.27 | 0.003 | 6,56 | 2.83 ; 10.28 | 0,0009 | 3,13 | -0.88 ; 7.14 | 0,12 |
| **Frontal lobe** | 0,022 | 0.00 ; 0.04 | **0.027** | 0,022 | 0.00 ; 0.04 | **0,02** | 0,017 | -0.00 ; 0.04 | 0,092 |
| Sex | -0,31 | -0.81 ; 0.18 | 0,21 | -0,33 | -0.80 ; 0.13 | 0,16 | -0,14 | -0.63 ; 0.35 | 0,57 |
| PMA at MRI | 0.021 | -0.07 ; 0.11 | 0,65 | -0,0067 | -0.09 ; 0.08 | 0,88 | 0,088 | -0.00 ; 0.18 | 0,062 |
| MRI Cohort | 0.048 | -0.35 ; 0.44 | 0,81 | 0,27 | -0.10 ; 0.64 | 0,15 | 0,31 | -0.09 ; 0.70 | 0,12 |
| SES | 0.093 | -0.00 ; 0.19 | 0.053 | 0,11 | 0.02 ; 0.20 | 0.015 | 0,084 | -0.01 ; 0.18 | 0,081 |
| | Adj R²: 0.12 | | | Adj R²: 0.17 | | | Adj R²: 0.15 | | |
| | p Model: 0.039 | | | p Model: 0.011 | | | p Model: 0.015 | | |
| Intercept | 6,63 | 2.61 ; 10.65 | 0.002 | 7,08 | 3.27 ; 10.90 | 0,0005 | 3,39 | -0.59 ; 7.37 | 0,094 |
| **Parietal lobe** | 0,027 | -0.00 ; 0.05 | 0.052 | 0,018 | -0.01 ; 0.04 | 0,18 | 0,022 | -0.01 ; 0.05 | 0,11 |
| Sex | -0,28 | -0.77 ; 0.22 | 0,27 | -0,23 | -0.71 ; 0.25 | 0,34 | -0,12 | -0.62 ; 0.37 | 0,61 |
| PMA at MRI | 0.023 | -0.07 ; 0.12 | 0,63 | 0,0047 | -0.08 ; 0.09 | 0,92 | 0,088 | -0.01 ; 0.18 | 0,064 |
| MRI Cohort | 0,14 | -0.28 ; 0.55 | 0,51 | 0,32 | -0.08 ; 0.72 | 0,11 | 0,38 | -0.03 ; 0.79 | 0,068 |
| SES | 0.094 | -0.00 ; 0.19 | 0.053 | 0,12 | 0.03 ; 0.22 | 0.012 | 0,084 | -0.01 ; 0.18 | 0,082 |
| | Adj R²: 0.097 | | | Adj R²: 0.11 | | | Adj R²: 0.15 | | |
| | p Model: 0.061 | | | p Model: 0.048 | | | p Model: 0.017 | | |
| Intercept | 7,2 | 3.19 ; 11.22 | 0,0007 | 7,5 | 3.75 ; 11.25 | 0,0002 | 3,86 | -0.09 ; 7.81 | 0.055 |
| **Occipital lobe** | 0,05 | -0.01 ; 0.11 | 0,077 | 0,046 | -0.01 ; 0.10 | 0,086 | 0,046 | -0.01 ; 0.10 | 0,1 |
| Sex | -0,22 | -0.70 ; 0.27 | 0,37 | -0,23 | -0.69 ; 0.23 | 0,32 | -0,093 | -0.57 ; 0.38 | 0,7 |
| PMA at MRI | 0.015 | -0.08 ; 0.11 | 0,77 | -0,011 | -0.10 ; 0.08 | 0,82 | 0,077 | -0.02 ; 0.17 | 0,12 |
| MRI Cohort | 0,0023 | -0.40 ; 0.40 | 0,99 | 0,22 | -0.16 ; 0.60 | 0,25 | 0,27 | -0.12 ; 0.67 | 0,18 |
| SES | 0.091 | -0.01 ; 0.19 | 0.068 | 0,12 | 0.02 ; 0.21 | 0.017 | 0,078 | -0.02 ; 0.17 | 0,11 |
| | Adj R²: 0.086 | | | Adj R²: 0.13 | | | Adj R²: 0.15 | | |
| | p Model: 0.078 | | | p Model: 0.031 | | | p Model: 0.016 | | |
| Intercept | 6,87 | 3.00 ; 10.75 | 0.0008 | 7,21 | 3.51 ; 10.91 | 0,0003 | 3,64 | -0.33 ; 7.60 | 0,072 |
| **Temporal lobe** | 0,044 | 0.01 ; 0.08 | **0.008** | 0,033 | 0.00 ; 0.06 | **0.035** | 0,024 | -0.01 ; 0.06 | 0,14 |
| Sex | -0,39 | -0.89 ; 0.10 | 0,12 | -0,33 | -0.81 ; 0.15 | 0,17 | -0,14 | -0.65 ; 0.37 | 0,59 |
| PMA at MRI | -0.00061 | -0.10 ; 0.09 | 0,99 | -0,016 | -0.11 ; 0.07 | 0,73 | 0,082 | -0.02 ; 0.18 | 0,096 |
| MRI Cohort | 0.046 | -0.34 ; 0.43 | 0,81 | 0,27 | -0.11 ; 0.64 | 0,16 | 0,3 | -0.09 ; 0.70 | 0,13 |
| SES | 0.09 | -0.00 ; 0.18 | 0.058 | 0,11 | 0.02 ; 0.20 | 0.017 | 0,085 | -0.01 ; 0.18 | 0,077 |
| | Adj R²: 0.15 | | | Adj R²: 0.15 | | | Adj R²: 0.14 | | |
| | p Model: 0.016 | | | p Model: 0.016 | | | p Model: 0.020 | | |
| Intercept | 7,98 | 4.00 ; 11.97 | 0,0002 | 8 | 4.19 ; 11.81 | < 0.0001 | 4,41 | 0.42 ; 8.41 | 0.031 |
| Cerebellum | 0,06 | 0.01 ; 0.11 | **0.017** | 0.041 | -0.01 ; 0.09 | 0.081 | 0.044 | -0.00 ; 0.09 | 0.077 |
| Sex | -0,24 | -0.70 ; 0.22 | 0.31 | -0,23 | -0.68 ; 0.23 | 0,33 | -0.083 | -0.55 ; 0.38 | 0,72 |
| PMA at MRI | -0,011 | -0.11 ; 0.09 | 0,83 | -0,02 | -0.12 ; 0.08 | 0,69 | 0.065 | -0.04 ; 0.17 | 0,2 |
| MRI Cohort | -0,052 | -0.45 ; 0.34 | 0,79 | 0,2 | -0.19 ; 0.58 | 0,31 | 0,23 | -0.16 ; 0.63 | 0,24 |
| SES | 0,11 | 0.01 ; 0.20 | 0.026 | 0,13 | 0.04 ; 0.22 | 0.006 | 0.093 | 0.00 ; 0.19 | 0.048 |
| | Adj R²: 0.13 | | | Adj R²: 0.13 | | | Adj R²: 0.16 | | |
| | $p$ Model: 0.028 | | | $p$ Model: 0.029 | | | $p$ Model: 0.013 | | |

## Supplementary Note

## Validation of the atlas prior based volumetry method

### Patient population

We validated two approaches for automatic brain volumetry: atlas prior based estimation (unpublished method used for testing the main hypothesis of this manuscript) and an expectation maximization based algorithm, implemented in the processing pipeline of the developing human connectome project dHCP (Makropoulos et al., 2014). Three different, overlapping subsets of the total study population were used for validation (Supplementary Table S7).

**Supplementary Table S7. Basic demographic data of the validation sub-populations.** Sub-set 3 is part of sub-set 2, and sub-set 2 is part of sub-set 1. CHD: congenital heart defect

| | Sub-set 1: Automatic volumetry validation dataset | Sub-set 2: Manual stereology of cerebellum, total brain volume and intracranial volume | Sub-set 3: Stereology incl. cortex and white matter volumes |
|---|---|---|---|
| Number of cases | 52 | 40 | 28 |
| Male/female | 34/18 | 28/12 | 17/11 |
| CHD / controls | 35/17 | 30/10 | 18/10 |
| Corrected gestational age at time of MRI (weeks, mean ± SD, range) | 41.4 ± 2.1, 37.6 – 46.2 | 41.2 ± 1.9, 37.6 – 45.1 | 41.3 ± 1.9, 38.1 – 45.1 |

### Image processing

The validation was carried out on the reconstructed 3D T2 images. First, non-brain tissue parts of the axial, coronal and sagittal T2-weighted images were removed by applying image masks generated from an age-specific neonatal neuroanatomical atlas. The masked, axial and coronal images were co-registered to the sagittal image using a mutual information based affine and non-linear registration as implemented in the Niftireg image registration package (Modat et al., 2010). Bias field correction of the images was performed by the N4ITK filter in the Slicer 3D software (Fedorov et al., 2012), and the three orthogonal resolution images were resampled to form a joint, high-resolution image. Examples for the reconstructed images and segmentations are given in the main manuscript.

### Automatic segmentation by non-linear transformation of anatomical priors

Anatomical priors from the ALBERTs[1] data set, which is gestational age-specific neonatal anatomical atlas (Gousias et al., 2012), were matched to the subjects reconstructed images using a linear affine registration (flirt in FSL) followed by a fast freeform non-linear deformation (reg_f3d command in the Niftireg software), based on the algorithm of Modal et al. 2010. The algorithm uses cubic B-spline to deform a source image in order to optimise an objective function based on the normalized mutual information and a penalty term. The penalty term used was the bending energy.

The parameters of the non-linear registration were the following:

- Grid spacing along the x,y,z directions: 8 * 8 * 8 mm

- Penalty term: bending energy=0.005

The anatomical masks of the supratentorial white matter, grey matter (cortex only) and lobar subdivisions of the supratentorial white matter (frontal, parietal, temporal and occipital lobe) were matched to the T2-weighted super-resolution images. A whole-brain mask was defined to include the diencephalon, cortical mantle, supratentorial white matter, but not the intra- and extra axial cerebrospinal fluid spaces or cerebellum. The patient-specific priors were thresholded at 50 % probability, and their volume was stored for the statistical evaluations.

### Validation using the automatic segmentation by dHCP pipeline

The dHCP structural pipeline was used to automatically segment the neonatal brain (Makropoulos et al., 2018). This pipeline is a fully automated processing pipeline for T1 and T2 weighted neonate brain MRI images. The pipeline performs cortical and sub-cortical volume segmentation, cortical surface extraction, and cortical surface inflation. The segmentation is based on an adaptive Expectation-Maximization algorithm (Makropoulos et al., 2014, 2016).

### Manual volumetry

Manual validation was performed by stereology using the EasyMeasure software package (Puddephat, 1999), following a method that has previously been described in detail elsewhere (Mayer et al., 2016). During stereology, a random grid is overlaid on the MR image, and the user assigns a label to each grid point that belongs to a given structure (Supplementary Figure S2).

---

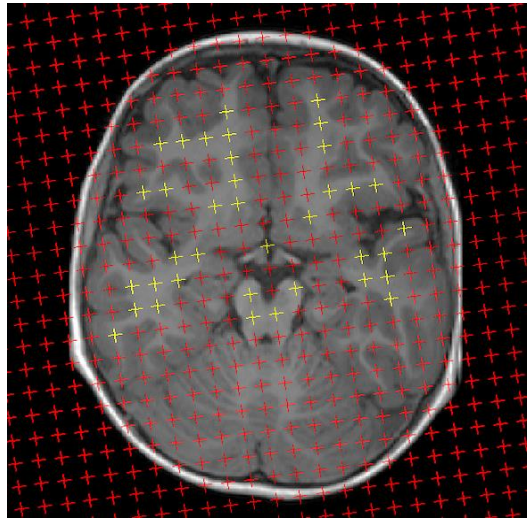

**Supplementary Figure S2. Manual volume estimation by stereology in the EasyMeasure software.** Grid points are displayed in red, overlaid on a T2-weighted neonatal MR image. (Illustration not actually used for measurement)

### *Correlations and overlaps of regional volumes across methods*

The agreement of the two automated volumetry approaches with ground truth manual stereology data was tested by calculating the correlation coefficient (Supplementary Table S8).

**Supplementary Table S8. Estimation of regional brain volumes and correlations with manual, ground truth stereology.**

| Region, method | Correlation coefficient (Pearson's r) | Significance (p) |
|---|---|---|
| Total brain volume, Atlas[1] | 0.930 | $4.56*10^{-18}$ |
| Cerebellum, Atlas[1] | 0.827 | $4.89*10^{-11}$ |
| Cortex, Atlas[1] | 0.881 | $6.27*10^{-10}$ |
| White matter, Atlas[1] | 0.675 | 0.000111 |
| Cerebellum, dHCP[2] | 0.914 | $4.53*10^{-16}$ |
| Cortex, dHCP[2] | 0.954 | $1.18*10^{-15}$ |
| White matter, dHCP[2] | 0.566 | 0.002 |

1: Atlas prior based approach, used during the main hypothesis test.

2: Method by Makropoulos et al.

As stereology does not result in binarized label maps, we were only able to calculate the spatial overlaps between the two automated approaches. Volumetric overlap was estimated as the Dice coefficient (%) of the subject-matched anatomical priors (binarized at 50%) and the transformed label maps of the dHCP pipeline. Table S2/3. summarizes the accuracy of the atlas prior based approach to the dHCP pipeline and manual stereology.

**Supplementary Table S9. Correlation of volume estimates and spatial overlaps with the atlas prior based approach.**

| Region, method | Correlation of volume estimates (Pearson's r) | Correlation of volume estimates, significance (p) | Volumetric overlap (Dice coefficient, %) |
|---|---|---|---|
| Cerebellum, dHCP[1] | 0.940 | $1.53*10^{-24}$ | 90.7 ± 2.2 |
| Cortex, dHCP[1] | 0.877 | $3.44*10^{-17}$ | 74.1 ± 2.6 |
| White matter, dHCP[1] | 0.745 | $3.67*10^{-10}$ | 80.4 ± 2.4 |
| Cerebellum, Manual | 0.856 | $3.55*10^{-12}$ | - |
| Cortex, Manual | 0.881 | $6.27*10^{-10}$ | - |
| White matter, Manual | 0.675 | 0.000111 | - |
| Frontal WM, dHCP[1] | 0.511 | 0.000108 | 82.7 ± 3.2 |
| Parietal WM, dHCP[1] | 0.547 | 0.000028 | 76.4 ± 6.3 |
| Temporal WM, dHCP[1] | 0.887 | $4.54*10^{-18}$ | 72.7 ± 2.8 |
| Occipital WM, dHCP[1] | 0.737 | $4.54*10^{-10}$ | 72.4 ± 3.4 |

1: Method by Makropoulos et al.

### *CHD / control volumetric ratios*

We tested whether the CHD to control ratio depends on the utilized segmentation or stereology approach (Figure S3).

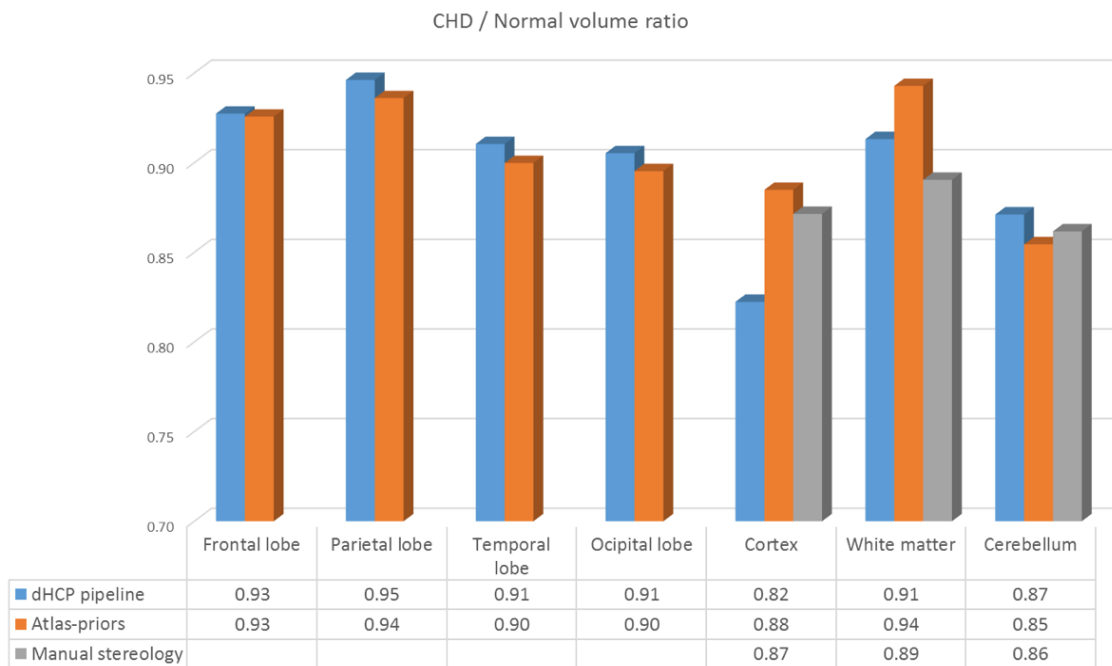

CHD / Normal volume ratio

| | Frontal lobe | Parietal lobe | Temporal lobe | Ocipital lobe | Cortex | White matter | Cerebellum |
|---|---|---|---|---|---|---|---|
| dHCP pipeline | 0.93 | 0.95 | 0.91 | 0.91 | 0.82 | 0.91 | 0.87 |
| Atlas-priors | 0.93 | 0.94 | 0.90 | 0.90 | 0.88 | 0.94 | 0.85 |
| Manual stereology | | | | | 0.87 | 0.89 | 0.86 |

**Supplementary Figure S3. Comparison of the CHD-to-normal brain volume ratios across methods.** Lobar volumes refer to lobar white matter volumes.

### *Validation of the atlas prior based approach - summary*

The anatomical atlas prior based segmentation and the dHCP pipeline gives comparable and faithful volumetry results for the cortex, total brain and cerebellar volume. The cortex and white matter volumes show larger variability, and less faithful estimation of the biologically relevant CHD/normal volume ratio. The volumes of lobar subdivisions show substantial differences between the methods (correlation coefficient 0.511-0.887), however, this could stem from the different anatomical definition of the deep white matter borders of the lobes in the two parcellation schemes. Despite the lower correlation of the lobar volumes, the corresponding CHD-to-normal ratios are remarkably similar between the two methods (difference: <2 %), while the cortex / white matter boundaries differed between these methods by a margin of 3-7%. The high variability of cortical volumes by both automated approaches limit the generalisability of the volumetric findings, and more reliable and reproducible results can be expected for lobar white matter, total brain volume and cerebellar volume estimates.

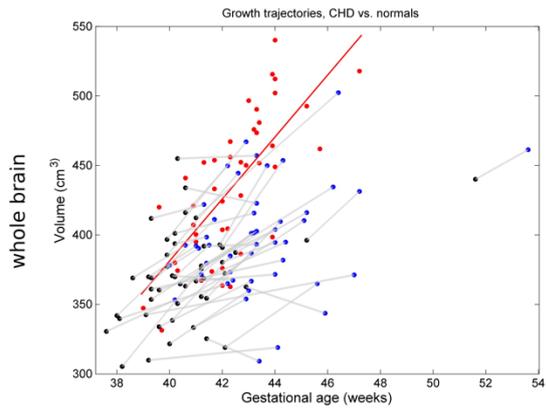
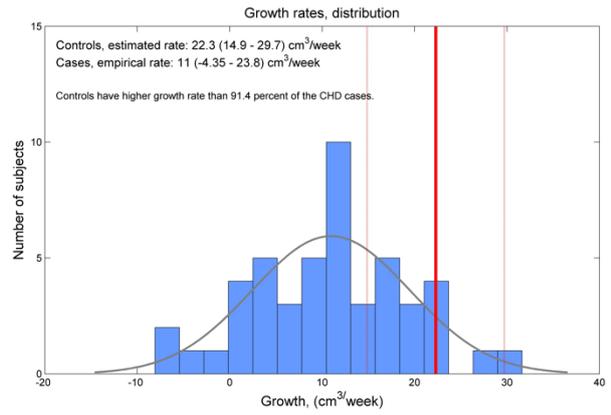

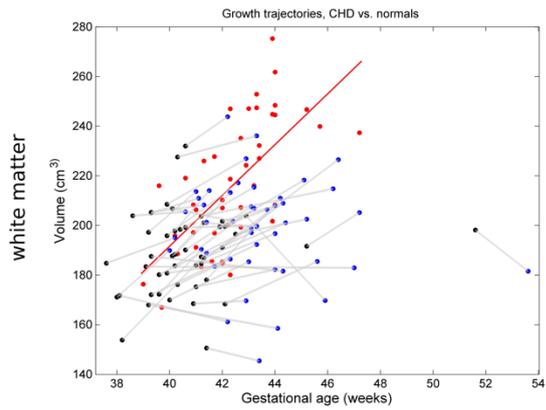
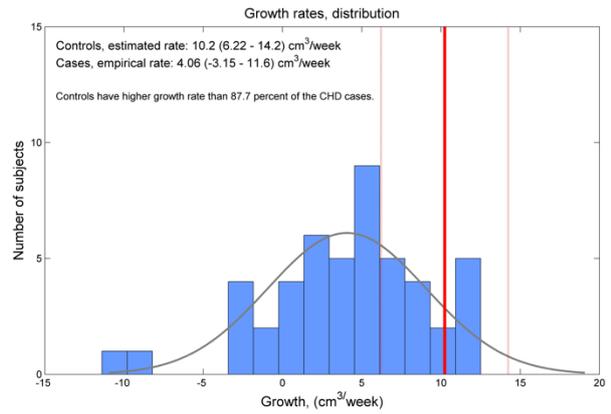

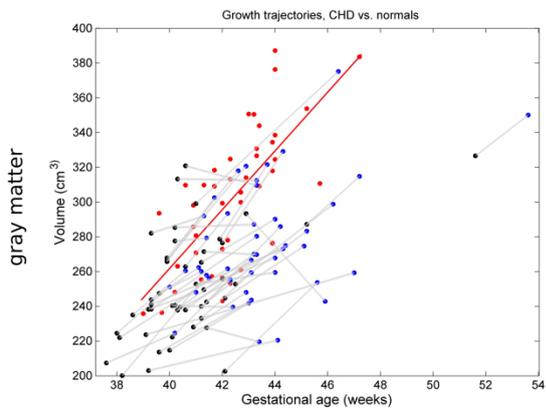
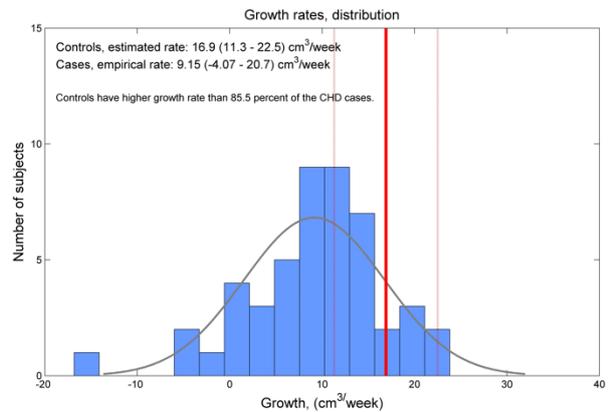

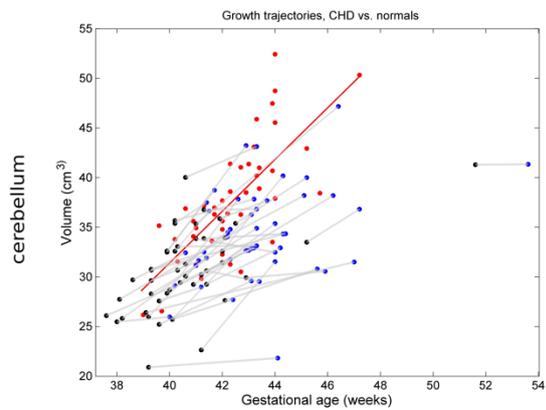
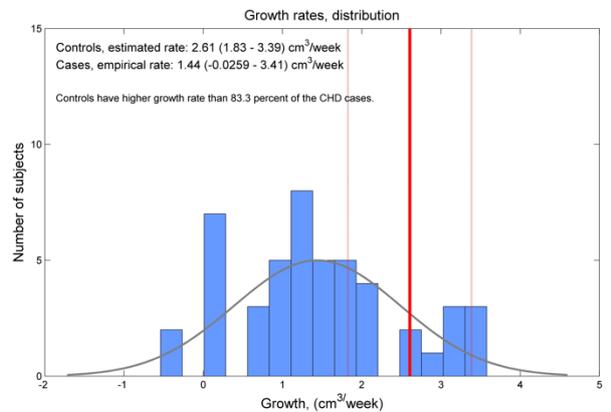

**Supplementary Figure S4. Growth trajectories of brain tissue subvolumes and cerebellum of CHD infants compared to controls.** Left images: linear regression plots. Red dots: brain volume of controls, Red line: cross-sectional brain growth of the controls. Black circles: preoperative brain volume of the CHD infants.

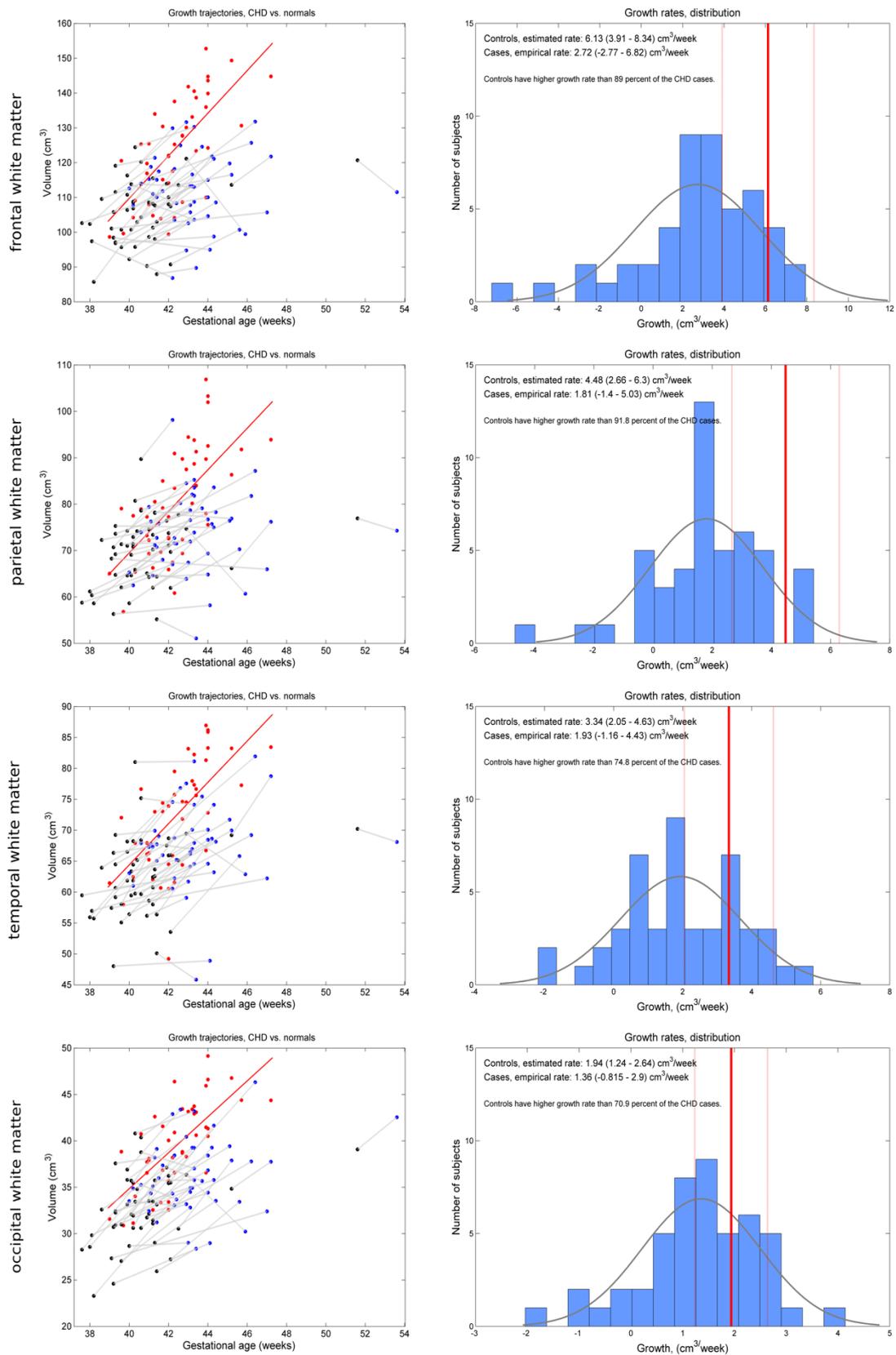

**Supplementary Figure S5. Growth trajectories of lobar white matter volumes of**

**CHD infants compared to controls.** Left images: linear regression plots. Red dots: brain volume of controls, Red line: cross-sectional brain growth of the controls. Black circles: preoperative brain volume of the CHD infants.